\documentclass[10pt,   showpacs, amsmath,amsfonts, amssymb,eqsecnum,  twocolumn, prd]{revtex4}
\usepackage{epsfig,eucal}

 \usepackage{amsfonts,amssymb}
\usepackage{amsmath}
\usepackage{youngtab}
\usepackage{graphicx}

\def\a{\alpha}
\def\b{\beta}
\def\g{\gamma}
\def\d{\delta}
\def\ep{\varepsilon}

\def\vp{\varphi}
\def\a{\alpha}
\def\b{\beta}

\def\g{\gamma}
\def\d{\delta}
\def\g{\gamma}

\def\J{{\mathcal{J}}}

 \def\H{{\mathcal{H}}}

\begin{document}
  \title{On skewon modification of  the light cone structure}
\author{Yakov Itin}
\email { itin@math.huji.ac.il}

\affiliation{Institute of Mathematics, The Hebrew University of
  Jerusalem \\ and Jerusalem College of Technology, Jerusalem,
  Israel. }

\pagestyle{myheadings}
\markboth{Yakov Itin} {Yakov Itin \qquad\qquad\qquad\qquad\qquad\qquad\qquad\qquad\qquad\qquad
{On skewon modification of light cone structure}}


 \begin{abstract}
Electromagnetic media with generic linear response provide a rich class of Lorentz violation models.  In the framework of a general covariant metric-free approach, we study electromagnetic wave propagation in these media.  
We define the notion of an optic tensor  and present  its unique canonical irreducible decomposition  into the principle and skewon parts. 
The skewon contribution to the Minkowski vacuum is a subject that does not arise in the ordinary models of Lorentz violation based on a modified Lagrangian. 
We derive the covector parametrization of the skewon optic tensor and discuss its $U(1)$-gauge symmetry.  We obtain several compact expressions for the contribution of the  principle and skewon optic tensor  to the dispersion relation. As an application  of the technique proposed here, we consider the case of a generic skewon tensor contributed to a simple metric-type principle part. 
Our main result: Every solution of the skewon modified Minkowski dispersion relation is necessary spacelike or null. It provides an extreme  violation of the Lorentz symmetry. The case of the antisymmetric skewon is  studied in detail and some new special cases (electric, magnetic, and degenerate) are discovered.  In the case of a skewon represented by a symmetric matrix, we observe a  parametric gap that has some similarity to the Higgs model.  
We worked out a set of specific examples that justify the generic properties of the skewon models and demonstrate  the different types of the Lorentz violation phenomena.   

\end{abstract}
\date{\today}
\maketitle

\section{Introduction }
As it is well known, the plain wave solution of the Maxwell electrodynamics system  yields the dispersion relation $\omega^2-{\bf k}^2=0$, which is the characteristic expression of Minkowski geometry. Propagation of electromagnetic waves on a curved manifold of GR is governed by the dispersion relation of a similar form $g^{ij}q_iq_j=0$, where $q_i=(\omega, {\bf k})$ is a wave covector. From this   basic expression of the pseudo-Riemann geometry one immediately derives that GR does not modify the point-wise light cone structure. 

On the other hand, modern field theories usually predict crucial modifications of the light cone structure expressed by an anisotropic dispersion relation, a birefringent vacuum and  a violation of local Lorentz and CPT invariance. Such theoretical phenomena emerge in loop  quantum gravity \cite{Gambini:1998it}, \cite{Bojowald:2004bb}, string theory \cite{Kostelecky:1991ak}, \cite{Kostelecky:1995qk} and the Very Special Relativity models \cite{Cohen:2006ky}, \cite{Gibbons:2007iu}. Since these models are still very far from their complete form and observational predictions, it is very important to have a phenomenological model that predicts the indicated optics phenomena.  

A well known achievement in this direction is the standard model extension (SME) construction due to Kostelecky and al. 
\cite{Kostelecky:2002hh}, \cite{Kostelecky:2007fx}, \cite{Kostelecky:2003fs}, \cite{Bluhm:2007bd}. In the electromagnetic sector of this construction, one starts with a covariant extension of the electromagnetic  Lagrangian involving a set of numerical parameters. the resulting field equations indeed yield solutions with breaking CPT symmetry and birefringence. However, the individual terms in the Lagrangian, apart from their covariance property, remain without a clear physical meaning. 

Another approach is based on a generalization of Riemann geometry to a generic Finsler geometry.  (See \cite{Perlick:2005hz}, \cite{Skakala:2008jp} for theoretical description, and   \cite{Lammerzahl:2008di}, \cite{Lammerzahl:2012kw}, and \cite{Itin:2014uia}  for observable effects analysis.) Since Finslerian geometry is anisotropic from  its very definition, the modified dispersion relation with CPT-violation and birefringence naturally emerges in this construction. What is unclear, however, is the physical reason for the Finslerian modification of space-time structure. Moreover, the Finsler geometry of the Lorentz type signature has serious  problems with its very definition,  that have not been resolved yet, see \cite{Pfeifer:2011tk}, \cite{Minguzzi:2014aua}. 

In this paper we study an alternative approach inspired    by  ideas from  solid state physics.  
It is well known that    the original formulation  of Maxwell's electrodynamics and GR  uses the medium analogy to a considerable extent.  It is rather natural to expect that  solid state analogies  may be useful for the modern  modifications of GR and other field theories.  
It should be taken into account that  classical GR is dealing  only with a relatively simple  second order metric tensor while  solid state physics (elasticity and electromagnetism) uses tensors of a higher order.  In modified field theory models,  higher order tensors  appear in a natural way. 

Electromagnetic wave propagation in media is a classical issue of mathematical physics. The standard presentation of this subject,  e.g.,  \cite{Zom}, \cite {Stratton}, \cite{LL}, \cite{Jack}, is essentially  non-relativistic (3-dimensional). Usually  textbooks restrict themselves to an  isotropic medium characterized by two scalar parameters $\ep$ and $\mu$ or to an anisotropic medium  described by two  symmetric $3\times 3$ matrices $\ep_{\a\b}$ and $\mu_{\a\b}$. 
It is clear, however, that a proper description of  wave propagation  must be presented in   terms of  4-dimensional tensorial  quantities defined on  space-time. This is mainly due to the  fact that the very  existence of  wave propagation should be  independent of a motion of an observer. 

Furthermore, the propagation of waves in an aniso\-tropic medium must be described by the medium parameters. It is not directly connected to the vacuum Minkowski metric. 
Thus the description of the electromagnetic wave phenomena must be independent of the metric structure of the space-time. It is well  known,   \cite{Kottler}, \cite{Cartan},  \cite{Dantzig},  \cite{Post}, that such a metric-free description of  classical electromagnetism is admissible. In this generic formalism, vacuum electrodynamics on a flat space-time and even on a curved space-time of general relativity is presented  as  specially simple cases.   Recent investigations  established  the metric-free electrodynamics as a logically complete construction based on the conservation laws for the electric charge, the magnetic flux, and the energy-momentum current. For details, see \cite{birkbook} and the references given therein. 

One of  the  valuable achievements of this construction is the proof that  the most general linear electromagnetic medium is characterized by a set of 36 independent parameters that form a 4-th order pseudo-tensor $\chi^{ijkl}$. This replaces the set of 12 material parameters given by  two 6-parametric sets, $\ep_{\a\b}$ and $\mu_{\a\b}$, in the classical representation. Due to the standard group theory  paradigm, the pseudo-tensor $\chi^{ijkl}$ is  decomposed into three independent pieces:
 (i) The {\it principal part} $ {}^{(1)}\chi^{ijkl}$ with 20 independent components; 
 (ii) The {\it skewon part} $ {}^{(2)}\chi^{ijkl}$ with 15  independent components;
  (iii) The {\it axion part} $ {}^{(3)}\chi^{ijkl}$ with 1 component.  
These pieces are unique, canonical, and irreducible under the action of the general linear group $GL(4,{\mathbb R})$. 
The principle part, $ {}^{(1)}\chi^{ijkl}$, comes as a covariant extension  of the classical set of 12  components of the 3-dimensional tensors $\ep_{\a\b}$ and $\mu_{\a\b}$.  The other two  parts, the skewon $ {}^{(2)}\chi^{ijkl}$ and  the axion $ {}^{(3)}\chi^{ijkl}$, do not have  classical analogs. 

The current paper is  devoted to the theoretical description of  wave propagation in  a generic linear response medium. 
Our aim is to understand the possible new physical phenomena coming from the additional electromagnetic parameters.  Recently there  has been   considerable progress in this area, \cite{Obukhov:2000nw}, \cite{Obukhov:2002xa}, \cite{Lammerzahl:2004ww}, \cite{Obukhov:2004zz}, \cite{L-S}, \cite{PostCon}, \cite{Itin:2005iv}, \cite{Itin:2007av}, \cite{Itin:2009aa}, \cite{Favaro:2010ys}, \cite{Dahl:2011he}, \cite{Itin:2012ann},  \cite{Perlick:2010ya}. Here we are interested mostly in the skewon effects on the light cone structure. It must be noted that skewon does not contribute to the electromagnetic Lagrangian, thus such effects cannot be appear in Kostelecky's construction nor in Finsler space schemes.

The organization of the paper is as follows: In Section 2, we  introduce our basic notations and present the integral Maxwell equations  and jump conditions in a metric-free form.   We discuss the linear constitutive relation. The decomposition of the constitutive tensor is based on the  technique Young diagrams. It proves that the decomposition is irreducible and unique (canonical). 
In Section 3, we give a generic definition of the wave-type solutions and present the  dispersion relation for these  solutions. 
The characteristic system is derived from the jump conditions. We show that it is  completely described by a Christoffel-type tensor $M^{ij}$ which is a close  analog of the acoustic Christoffel 3D-matrix. We provide a decomposition of  $M^{ij}$  into two independent optic tensors $P^{ij}$ and $Q^{ij}$ defined by the principal and the skewon part correspondingly. Moreover, the show that skewon part $Q^{ij}$ is of the form $q_iY_j-q_jY_i$. Thus the  covector $Y_i$ completely describes skewon effects. 
In Section 4, the dispersion relation in terms of the tensors $P^{ij}$ and $Q^{ij}$ is derived. We are able to express the whole contribution of the skewon to  the  light cone structure in one compact term, which thus allows a qualitative analysis. 
Section 5 is devoted to an analysis of the skewon effects on the wave propagation in a (pseudo)Riemannian vacuum. In particular, we prove that an arbitrary skewon extends the light cone and thus yields superluminal wave propagation. 
 In section 6  we study the special case of an antisymmetric skewon. We reinstate the known fact that the Fresnel surface separates into two surfaces. We prove that  both surfaces have the  Lorentz signature and can thus  serve as light cones. Moreover we show that for an antisymmetric skewon these cones always intersect along two optic axes. The variety of mutual arrangements is still rich. In particular, we identify the electric, magnetic and degenerate types of antisymmetric skewon. 
Section 7 deals with the case of the symmetric skewon. We show that in this case    wave propagation does not exist for small values of the skewon parameters and reappears only for  sufficient big values of the skewon. This behavior is similar to the one exhibited by the Higgs model in  particle physics. We describe different parametrizations of the symmetric skewon that can be characterized by the rank of a corresponding matrix. Explicit examples are provided. 
In the concluding section, we discuss our results and their possible extensions. 
Some technical details are delayed to the  Appendix.

\section{A covariant description of electromagnetism in anisotropic media}
\subsection{Maxwell's system}
Our goal is to study  wave propagation in an  electromagnetic medium with a generic   local linear response. 
Vacuum  can be considered as a specific type  of a unique "medium" completely characterized by the metric tensor.  
In  contrast, different  types of real anisotropic media, be  characterized  each by their own sets of parameters, may  present simultaneously in the same problem. We must, for instance, often  deal with two media separated by a thin boundary. In  phenomenological models, such a boundary is  described by a smooth 2-dimensional surface which either stationary or moving  in   3-dimensional space. In the 4-dimensional space-time representation, the stationary or moving boundaries are described in the same fashion as  smooth 3-dimensional hypersurfaces.  In such a situation, some components of the field necessary  have a jump through the boundary surface. Thus, the electromagnetic field  in media  is,  in general, non-differentiable and the ordinarily used partial differential field equations  are not directly applicable. Quite similarly, the non-differentiable fields emerge in geometrical optics  at the wave fronts even in the vacuum case.

A well  known  precise mathematical  way to deal with such non-differentiable fields is to consider the integral field equations instead of the differential ones. The most natural objects for  integration on a differential manifold are  differential forms.  From  their very definition, the   differential forms are correlated to  the dimension of the integration domain, thus they provide a covariant formulation of physics laws.  For instance, in terms of differential forms,  the integral Maxwell equations are given in a  compact covariant form 
\begin{equation}\label{max-1}
 \int_{M_2}F=0\,,\qquad \int_{N_2}\H= \int_{N_3}\J\,.
\end{equation}
Here $F$ is the even (untwisted) 2-form of {\it field strength}, while $\H$ is the odd (twisted) 2-form of  {\it excitation}, $\J$ is the odd 3-form of  {\it electric current}.   
 The integration domain  $M_2$ is a smooth closed 2-dimensional surface, while $N_3$ is a smooth 3-dimensional oriented hyper-surface with a smooth boundary $N_2=\partial N_3$. The case of  non-smooth domains is sometimes also useful for physics applications. There an extension of the formulation given in Eqs.(\ref{max-1}) can be provided by a limiting procedure. In this paper, we restrict ourselves to the case of smooth integration domains. 

 Eqs.(\ref{max-1}) have a clear physical interpretation. In particular, the first equation  represents the  conservation law of the magnetic flux. 
The second equation of (\ref{max-1})  may be viewed as   a consequence of the  electric current conservation law. For the details of  such an interpretation, see  \cite{birkbook}.
\subsection{Covariant jump conditions and wavefront}
In order to derive the wave-front conditions, we apply the covariant jump conditions, \cite{Itin:2012ann}, that are a consequence of the integral Maxwell equations (\ref{max-1}).  Consider  an arbitrary smooth non-degenerate hypersurface $\Sigma$ in  space-time. Let its implicit description  be given by the equation 
  \begin{equation}\label{jump-0}
\vp(x^i)=0\,.
\end{equation}
Thus, at an arbitrary  point of $\Sigma$, the covector $q_i=\partial  \vp/\partial x^i$ is assumed to be well-defined and non-zero. Here and in the sequel, the Roman indices take values in the range $\{0,1,2,3\}.$ 
We define two limit values
\begin{eqnarray}
F_{\pm}&=&\lim_{\ep\to 0}F(x^i)\big|_{\vp(x^i)=\pm\ep}\,.
\end{eqnarray}
The jump $\big[F\big]$ of the field strength $F$  at the hypersurface $\Sigma$ is defined as  the difference of these two  values
\begin{equation}\label{bound-form3}
 \big[F\big] =F_{+}-F_{-}\,.
\end{equation}
A similar definition holds  for the jump $\big[\H\big]$ of the excitation $\H$.

In the source-free case, the  {\it jump conditions} are given by the following covariant and metric-free equations: 
\begin{equation}\label{bound-form}
 \big[F\big] \wedge d \vp =0\,,\qquad  \big[\H\big] \wedge d\vp =0\,.
\end{equation}
These equations  are straightforward consequences of the integral field equations (\ref{max-1}) provided  the sources do not enter the region. 
The  tensorial form of Eqs.(\ref{bound-form}) read as
\begin{equation}\label{bound-form4}
\ep^{ijkl}\big[F_{jk}\big] \vp_{,l}=0\,,\qquad \big[\H^{ij}\big]\vp_{,j}=0\,.
\end{equation}
Here $\ep^{ijkl}$ is the permutation pseudotensor normalized by $\ep_{0123}=1$. 

Since the jump conditions (\ref{bound-form},\ref{bound-form4}) are metric-free, they can be used for a wide class of applications. The hyper-surface $\Sigma$ may represent:
\begin{description}
\item[(1)] {\it A Cauchy initial surface} that is ordinarily required to be {\it time-like. } 

\item[(2)]    {\it A boundary surface} that separates  two different media. It is  ordinarily represented by  a {\it space-like} hypersurface. 

\item[(3)]   {\it A wavefront} that  emerges  in  4-dimensional geometrical optics as a {\it light-like} hypersurface.   
\end{description}

The covariant boundary surface problem was studied in \cite{Lindell} and \cite{Itin:2012ann}. In particular, 
when the space-time decomposition of the fields $F$ and $\H$ is applied and the fields are assumed to be independent of the time coordinate,   
Eqs.(\ref{bound-form}) yield the standard boundary conditions for the 3-dimensional fields $E, B, D$ and $H$. In the current paper, we are dealing with   electromagnetic wave propagation, i.e., with the wavefront problems. 

\subsection{Constitutive pseudotensor}
Eqs.(\ref{max-1}) present only  the formal structure of electromagnetism. They are endowed with a physical content  only when a constitutive relation between the fields $\H$ and $F$ is  postulated. In general, such a relation can be  non-linear and even non-local (as in  ferromagnetics). In this paper, we consider the simplest but  rather widespread case of a {\it linear,  local   constitutive relation: }
  \begin{equation}\label{lin-1}
\H^{ij}=\frac 12 \chi^{ijkl}F_{kl}\,.
\end{equation}
Due to its definition, the {\it constitutive pseudotensor} $\chi^{ijkl}$ possess the symmetries
  \begin{equation}\label{lin-1x}
\chi^{ijkl}=-\chi^{jikl}=-\chi^{ijlk}\,.
\end{equation}
Hence, in 4-dimensional space, $\chi^{ijkl}$ has   36 independent components. In general, the  constitutive pseudotensor is a field that   depends of the point on the manifold. 

A useful way to deal with  a  multi-component tensor, such as $\chi^{ijkl}$, is to  decompose it into the sum of simpler independent  sub-tensors with fewer components. Naturally, these  parts   must themselves be tensors of the same rank. Moreover,  it is preferable for this decomposition to be a {\it unique and irreducible.} In this case, to the partial sub-tensors might be endowed with independent physical meaning.  
So, we are are looking for a decomposition of a 4-rank tensor  defined on  4-dimensional space-time.  Moreover, we must take into account that $\chi^{ijkl}$ is not a general 4-th order tensor, it carries its original symmetries (\ref{lin-1x}). 
In other words, the tensor $\chi^{ijkl}$ itself constitutes a subspace of a generic tensor space. Consequently we need a decomposition of an invariant  subspace into the  direct sum of its invariant (sub)subspaces. A common  way to derive such a decomposition is to apply the Young diagram technique. A brief account of this subject is given in \cite{Baekler:2014kha}.  Note that a generic 4-rank tensor (without additional symmetries)  is canonically decomposed into five sub-tensors. An important fact that these partial tensors are  yet reducible. Their successive  decomposition  is irreducible but not canonical and thus non-unique. The situation is changed drastically when we decompose a tensor with additional symmetries. 
First,  the Littlewood-Richardson rule restricts the number of the relevant Young's diagrams. In our case, there are only three different types of  diagrams 
\begin{eqnarray}\label{tab0}
\Yvcentermath1\yng(1,1)\,\,\otimes\,\,\yng(1,1)\,\, =\,\,\Yvcentermath1 \,\yng(2,2)\Yvcentermath1\,\,\oplus\,\,\yng(2,1,1)\,\,\oplus\,\,\yng(1,1,1,1)\,.
\end{eqnarray}
The dimensions of the subspaces depicted in the diagrams (\ref{tab0}) are 20, 15, and 1, correspondingly. It  proves that there is a unique decomposition of the set of the components
\begin{equation}\label{lin-3x}
36=20+15+1\,.
\end{equation}
Inasmuch as  there are no invariant subspaces of smaller dimensions, the   canonical decomposition of the constitutive pseudotensor into three pieces is unique and irreducible. 

Following \cite{birkbook},  we denote the decomposition depicted in  (\ref{tab0}) as 
\begin{equation}\label{lin-3}
\chi^{ijkl}={}^{\tt (1)}\chi^{ijkl}+{}^{\tt (2)}\chi^{ijkl}+{}^{\tt
(3)}\chi^{ijkl}\,.
\end{equation}

Here the {\it axion part}  ${}^{\tt(3)}\chi^{ijkl}$ is represented by the third diagram of (\ref{tab0}). It represents the complete antisymmetrization of $\chi^{ijkl}$ in all its four indices.  This pseudotensor has only one independent component. Consequently it can be represented by a pseudoscalar $\alpha$,
\begin{equation}\label{lin-4}
{}^{\tt (3)}\chi^{ijkl}=\chi^{[ijkl]}=\alpha\varepsilon^{ijkl}\,.
 \end{equation}
Such  pseudoscalar is a rather popular object in particle physics. Note that in the current setting axion appears rather naturally and is not artificially added. 

The {\it skewon part} ${}^{\tt (2)}\chi^{ijkl}$ of 15 independent components corresponds to the middle diagram of (\ref{tab0}). It is expressed as 
\begin{equation}\label{lin-5}   
{}^{\tt (2)}\chi^{ijkl}=\frac
12\left(\chi^{ijkl}-\chi^{klij}\right)\,.
 \end{equation}

 The {\it principal part} of 20 independent components corresponds to the first diagram of (\ref{tab0}). It is expressed as
\begin{equation}\label{lin-7}  
{}^{\tt (1)}\chi^{ijkl} =\frac 16\Big(2\chi^{ijkl}+2\chi^{klij}-\chi^{iklj}-\chi^{ljik}-\chi^{iljk}-\chi^{jkil}\Big)\,.
\end{equation}
It is now straightforward to check   that the sum of the terms (\ref{lin-4}), (\ref{lin-5}), and    (\ref{lin-7}) is equal to the initial tensor $\chi^{ijkl}$. A derivation of the expressions (\ref{lin-4}), (\ref{lin-5}), and    (\ref{lin-7}) can be done by the Young diagram technique. The result is equivalent to the derivation based on the $6\times 6$ representation \cite{birkbook}. Group theoretical  considerations provide a proof  that this decomposition is canonical, unique and irreducible under the action of the general linear group.

\section{Optics tensors}
Our goal is to study  the  propagation of electromagnetic waves in a medium with a generic constitutive pseudotensor $\chi^{ijkl}$. Particularly, we are interested in the contributions of the irreducible parts ${}^{\tt (1)}\chi^{ijkl}, {}^{\tt (2)}\chi^{ijkl}$ and ${}^{\tt (3)}\chi^{ijkl}$ to the shape of  the Fresnel surface  and to the corresponding light cone structure. In order to have a description that is valid for an arbitrary observer, we are looking for a covariant formulation. 
\subsection{Generalized wavefront and characteristic system}
We need  a {\it covariant} description of the electromagnetic wave in media.  Usually a wave is represented by a plane wave ansatz or by its generalization in series, see \cite{perlick-book}, \cite{Perlick:2010ya} for a modern review of this non-covariant approach. Instead, we will apply the following covariant description  that is similar to what is given in \cite{Lichn} and\cite{K-K}.

\vspace{0.2 cm}

{\bf Definition 1:}    {\it A generalized electromagnetic wave} is a set of the solutions,  $F$ and $  {\cal H}$,  of Maxwell's system (\ref{max-1})that are non-zero on one side of some hypresurface $\vp(x^i)=0$ and zero on its other side. 

With this definition, the generic jump conditions (\ref{bound-form})  take  the form of the {\it wavefront conditions}
\begin{equation}\label{int6a}
 F \wedge d\vp=0\,, \qquad  {\cal H} \wedge d\vp=0\,.
\end{equation}
In tensorial representation with the wave covector defined as $q_i=\partial \vp/\partial x^i$,  they read
\begin{equation}\label{int6b}
\ep^{ijkl}F_{jk}q_{l}=0\,,\qquad \H^{ij}q_{j}=0\,.
\end{equation}
The  hypersurface $\vp(x^i)=0$ appearing in Definition 1 is referred to as a {\it wavefront}.  


An   explicit expression of the wavefront  is determined from the  uniqueness and existence conditions for the solutions of (\ref{int6b}).  
When the linear constitutive relation (\ref{lin-1}) 
 is substituted here  we obtain 
\begin{equation}\label{char-1}  
\ep^{ijkl}q_jF_{kl} =0\,,\qquad\qquad \chi^{ijkl}q_jF_{kl} =0\,.
\end{equation}
This is linear system  of 8 equations for 6 independent variables, the components  of the tensor $F_{kl}$. However it  is   only formally overdetermined.  Indeed, there are two  linear identities between the equations. When the left hand sides of  (\ref{char-1}) are  multiplied by $q_i$ they vanish identically due to the symmetries (\ref{lin-1x}). 

How can we deal with this formally overdetermined system? Often one  substitutes the constraints into (\ref{char-1}) and end up with a well defined system of 6 independent equations of 6 independent variables. 
Note that  the field $F_{kl}$ is a {\it measurable quantity}, and is thus certain kind unique.  Therefore an  {\it  existence  and uniqueness theorem}  must hold: 
(i)  For {\it existence},  the $6\times 6$ determinant of the matrix of the coefficients must be equal to zero. 
(ii) For {\it uniqueness}, this determinant  must be non-trivial. Thus one comes to the $6\times 6$ determinant and correspondingly to a characteristic equation that is of the six order in $q$. 
This straightforward procedure explicitly violates  the covariance of the system. So a more precise question is: How can  we deal with the overdetermined system (\ref{char-1}) without violating  its covariance? 

In order to have a covariant derivation, we start with the first  equation of (\ref{char-1}). 
Its most general solution is of the form 
\begin{equation}\label{char-2}  
F_{kl} =\frac 12 (a_kq_l-a_lq_k)\,
\end{equation}
with an arbitrary covector $a_k$. In fact, $a_k$ can be viewed as an algebraic analog of the standard electromagnetic potential. 
Substituting (\ref{char-2}) into the second equation of (\ref{char-1}) we obtain the {\it characteristic system}
\begin{equation}\label{char-3}  
{M^{ik}a_k=0\,,}
\end{equation}
where the {\it characteristic matrix} is 
\begin{equation}\label{char-3x}  
M^{ik}=\chi^{ijkl}q_jq_l\,.
\end{equation}
Eq.(\ref{char-3}) is a linear system of 4 covariant equations for the 4 components of the covector $a_k$. Now we face  a new problem: Since the relations 
\begin{equation}\label{char-3x-cc} 
M^{ik}q_k=0\qquad {\rm and }\qquad M^{ik}q_i=0
\end{equation}
 hold identically for the matrix defined in (\ref{char-3x}), they can be treated as  linear relations between the rows and columns of the matrix $M_{ik}$. Consequently the system  (\ref{char-3}) is {\it undetermined.}
This was to be  expected since, in contrast to $F_{ij}$, the potential $a_k$ is an  unmeasurable quantity.  Consequently the solutions of (\ref{char-3})  cannot be unique. 
As shown in \cite{Itin:2009aa}, the relations  (\ref{char-3x-cc}) correspond to the physical principles --   gauge invariance and   charge conservation of our system.  
Thus, in order to preserve the gauge invariance and the covariance of the characteristic system, we must proceed with an undetermined characteristic system and with a singular characteristic matrix $M^{ik}$. 

\subsection{Two optic tensors} 

The electromagnetic wave propagation  is now described  by a linear characteristic system $M^{ik}q_k=0$.  In particular, it is completely determined  by the matrix $M^{ik}$.  
 This matrix is an analog of the {\it acoustic} (Christoffel) $3\times 3$ matrix which is used to describe  the acoustic wave propagation in linear elasticity. We will refer to $M^{ik}$ as the {\it optic tensor}. 

Substituting the irreducible decomposition (\ref{lin-3})  into 
Eq.(\ref{char-3x}), we obtain 
\begin{equation}\label{chis-1}  
M^{ik}={}^{(1)}\chi^{ijkl}q_jq_l+{}^{(2)}\chi^{ijkl}q_jq_l
+{}^{(3)}\chi^{ijkl}q_jq_l\,.
\end{equation}
The last term here is a contraction of symmetric and  skew-symmetric matrices, thus it vanishes identically 
\begin{equation}\label{chis-2}  
{}^{(3)}\chi^{ijkl}q_jq_l=\a\ep^{ijkl}q_jq_l\equiv 0\,.
\end{equation}
Consequently the  axion part  ${}^{(3)}\chi^{ijkl}$ does  not contribute to the wave propagation. This fact does not  contradicts   the known results about  axion modification of the dispersion relation \cite{Carroll:1989vb}. In fact, the axion contribution emerges only in the higher order approximation \cite{Itin:2004za}, \cite{Itin:2007wz}, \cite{Itin:2007cv}. In this paper we restrict ourselves to the standard  geometric optics approximation that does not take the axion contributions into account.

The two other terms on the right hand side of Eq.(\ref{chis-1}) are  in general non-zero.  
Consequently,  the optic matrix is irreducibly decomposed into the sum of two terms 
\begin{equation}\label{chis-4}  
M^{ik}=P^{ik}+ Q^{ik}\,.
\end{equation}
Here, the {\it principle optic tensor } $P^{ik}$ and the {\it skewon optic tensor }  $Q^{ik}$ are defined as 
\begin{equation}\label{chis-3}  
P^{ik}={}^{(1)}\chi^{ijkl}q_jq_l\,,\qquad Q^{ik}={}^{(2)}\chi^{ijkl}q_jq_l\,.
\end{equation}
Let us list some basic properties of these two tensors:

\begin{itemize}
\item[(1)] { \it  Symmetry:}
The { principle optic tensor } $P^{ik}$ is symmetric while the { skewon optic tensor }  $Q^{ik}$ is antisymmetric:  
\begin{equation}\label{chis-4-cc}  
P^{ik}=P^{ki}\,,\qquad  Q^{ik}=-Q^{ki}\,.
\end{equation}
\item[(2)]  {\it Linear relations:} Since the partial pseudo-tensors ${}^{(1)}\chi^{ijkl}$ and ${}^{(2)}\chi^{ijkl}$ preserve the symmetries of the original pseudo-tensor  $\chi^{ijkl}$, the linear relations of the type (\ref{char-3x-cc}) hold  also for the matrices  $P^{ik}$ and  $Q^{ik}$, i.e.,
\begin{equation}\label{chis-5}  
P^{ik}q_k=0\,,\qquad {\rm and}\qquad Q^{ik}q_k=0\,.
\end{equation} 
\item[(3)]  {\it Determinants:} Due to the linear relations between the columns of the matrices, we have, 
\begin{equation}\label{chis-5a}  
{\rm det}(P)\equiv 0\,,\qquad {\rm and}\qquad {\rm det}(Q)\equiv 0\,.
\end{equation} 
It is in  addition to the relation ${\rm det}(M)\equiv 0$.  
\item[(4)]  {\it Adjoint of the skewon optics tensor:} 
In order to have a non-trivial (non-zero) expression for the adjoint matrix,  the rank of the original $4\times 4$-matrix must be equal to 3. But  the rank of an arbitrary antisymmetric matrix is even.
Thus  our skewon matrix satisfies 
\begin{equation}\label{chis-5x}
  {\rm adj}(Q)\equiv 0\,.
\end{equation} 
\end{itemize}
We will see in the sequel,  that this fact yields that the skewon part alone does not provide  a non-trivial  dispersion relation. 
Thus skewon  can  serve only as a supplement to the principle part.  

\subsection{Skewon optics covector}
Since the skewon part of the constitutive tensor contributes to the wave propagation only via the antisymmetric tensor $Q_{ij}$, it has   a simpler representation, which we now derive 

We start with the relation 
\begin{equation}\label{chis-5xx}  
Q^{ij}q_j=0\,.
\end{equation} 
It is convenient to define  an auxiliary tensor
\begin{equation}\label{chis-6}  
{\check  Q}_{ij}=\frac 12\ep_{ijmk}Q^{mk}\,,\quad {\rm thus}\quad Q^{pq}=-\frac 12 \ep^{ijpq}{\check  Q}_{ij}\,.
\end{equation}
In this notations, Eq.(\ref{chis-5xx}) takes the form
  \begin{equation}\label{chis-8}  
 \ep^{ijpq}{\check  Q}_{ij}q_p=0\,.
\end{equation}
This equation is somewhat simpler  to deal with than (\ref{chis-5xx}) because it has  exactly the  same form as the first equation of the Maxwell system (\ref{char-1}). Hence the most general solution of (\ref{chis-8}) can be written in the form of  (\ref{char-2}),
 \begin{equation}\label{chis-9}  
{\check  Q}_{ij} =Y_iq_j-Y_jq_i\,
\end{equation}
with an arbitrary covector $Y_i$. We will refer to  $Y_i$    as the {\it skewon optic covector.} 
Let us list its basic properties: 
\begin{itemize}
\item[(1)]  Since ${\check  Q}_{ij}$ is  quadratic in $q_i$, the components of $Y_i$ are the {\it first order homogeneous functions} of  $q_i$;
\item[(2)]  $Y_i$ is a {\it covector density}, because ${\check  Q}_{ij}$ is a tensor density;
\item[(3)]  Due to (\ref{chis-9}), $Y_i$ is defined  only up to an arbitrary  addition of the wave covector $q_i$. So we have here a type of a {\it gauge symmetry} which is similar to the ordinary gauge symmetry of the Maxwell system;
\item[(4)]  An  additional {\it gauge fixing condition} on  $Y_i$  can be applied. 
\end{itemize}
Substituting  (\ref{chis-9})  into  Eq.(\ref{chis-6}) we obtain
\begin{equation}\label{chis-10}  
Q^{ij}=-\frac 12 \ep^{ijrs}(Y_rq_s-Y_sq_r)\,.
\end{equation}
Using  the skew-symmetry, this expression  can be rewritten finally as 
\begin{equation}\label{chis-11}  
Q^{ij}=\ep^{ijkl}q_kY_l\,.
\end{equation}
\subsection{Skewon in matrix representation}
Due to the gauge symmetry, the covector $Y_i$ can be expressed as 
\begin{equation}\label{mat-0}  
Y_i=S_i{}^jq_j+\theta q_i\,
\end{equation}
with an arbitrary scalar $\theta$ that represents the gauge freedom.  
We observe that in this representation  the tensor $S_i{}^j$ and the scalar $\theta$ must be  zero order homogeneous functions of the wave covector $q_k$. Without loss of generality we can assume the  tensor $S_i{}^j$ to be traceless, $S_i{}^i=0$, because its trace can be absorbed into the scalar $\theta$.
Like a traceless quadratic matrix, the skewon has  15 independent components. 
A representation of the skewon by a traceless mixed tensor was derived in \cite{birkbook},
\begin{equation}\label{mat-1}  
^{(2)}\chi^{ijkl}=\ep^{ijmk}S_m{}^l-\ep^{ijml}S_m{}^k\,.
\end{equation}
 Let us check that this expression  meets exactly our definition (\ref{mat-0}).
When we substitute (\ref{mat-1}) into Eq.(\ref{chis-3}), the skewon optic tensor becomes  
\begin{equation}\label{mat-2}  
Q^{ik}=^{(2)}\chi^{ijkl}q_jq_l=
\ep^{ijmk}S_m{}^lq_jq_l\,.
\end{equation}
Rearranging the indices we can rewrite this as 
\begin{equation}\label{mat-3}  
Q^{ij}=\ep^{ijkl}q_k\left(S_l{}^mq_m\right)\,.
\end{equation}
Compare Eq.(\ref{mat-3}) with the definition of the skewon covector  (\ref{chis-11}).  In the brackets of (\ref{mat-3}), we recognize (up to the $\theta$-factor) the covector $Y_i$ as it appears in Eq.(\ref{mat-0}). 
Thus the tensor $S_i{}^j$ turns out to be independent of $q$.  Alternatively, the scalar $\theta$ remains an arbitrary  zero order homogeneous functions of  $q_k$. 
  Different choices of  $\theta$ represent different gauge conditions. In the sequel, we will see that  the Lorenz-type gauge condition for the covector $Y_i$ requires a non-zero and even non-regular scalar function $\theta$. 
\section{Dispersion relation}
\subsection{Covariant dispersion relation}
  In order to have a non-trivial solution $a_k$ of the characteristic system
\begin{equation}\label{disp-1}  
M^{ik}(q)a_k=0\,.  
\end{equation}
 the components of the wave covector $q_i$ must satisfy certain consistent relation. This {\it dispersion relation}  is a polynomial  algebraic equation with the coefficients depending on the media parameters. 
Usually  an equation of the form (\ref{disp-1}) is consistent (has non-trivial solutions) when the relation ${\rm det} (M)=0$ holds. The electromagnetic system, however, is a singular one, so its determinant vanishes  identically. This fact is due to {\it the gauge invariance of the system.}  Indeed, $a_l\sim q_l$ is a  solution of Eq.(\ref{disp-1}) that does not contribute to the electromagnetic field strength, so it is unphysical. 
Hence, the  system (\ref{disp-1})  accepts  a
physical meaning, only if it has at least {\it two linearly independent solutions} -- one for gauge freedom and one for physics. We recall a known fact from linear algebra: A  linear system
has two (or more) linearly independent solutions if and only if
{ the rank of its matrix $M^{ij}$ is of 2 (or less)}. It means that the {\it adjoint} of the  matrix $M^{ij}$  must be  equal to zero. Recall that the adjoint matrix  is constructed from the cofactors of $M^{ij}$. 

Thus, in electromagnetism, as well as in an arbitrary $U(1)$-gauge invariant system,  existence of physically meaningful solutions of a system $M^{ik}(q)a_k=0$ requires
\begin{equation}\label{disp-2}  
{\rm adj}(M)=0\,.
\end{equation}
 The standard expression of the 4-th order adjoint matrix $A_{ij}={\rm adj}(M)$ is of the form 
\begin{equation}\label{disp-3-cc}  
A_{ij}=\frac 1{3!}\ep_{ii_1i_2i_3}\ep_{jj_1j_2j_3}M^{i_1j_1}M^{i_2j_2}
M^{i_3j_3} \,.
\end{equation}
Consequently the dispersion relation reads 
\begin{equation}\label{disp-3}  
 \ep_{ii_1i_2i_3}\ep_{jj_1j_2j_3}M^{i_1j_1}M^{i_2j_2}
M^{i_3j_3}=0 \,.
\end{equation}
The condition (\ref{disp-3}) is somewhat unusual because it is given  in a matrix form.  However, we have here only one independent condition. Indeed,  we can observe the following algebraic fact, see \cite{Itin:2009aa} for a proof.
 Let  an $n\!\times\! n$ matrix $M^{ij}$  satisfies 
   $M^{ij}q_i=0$ and $ M^{ij}q_j=0$ for some   $n$-covector $q_i\ne 0$. 
Then its  adjoint  matrix $A_{ij}={\rm adj}(M)$ is proportional to the tensor product of the covectors $q_i$,
    \begin{equation}\label{Adj2}
A_{ij}=\lambda(q)q_iq_j\,,
\end{equation}
where $\lambda(q)$ is a  polynomial of $q$.

\vspace{0.2 cm}
Thus (\ref{disp-3}) reads as 
    \begin{equation}\label{Adj2x}
\lambda(q)q_iq_j=0\,.
\end{equation}
Since $q_i$ is non-zero, the dispersion relation takes the ordinary scalar form  
 \begin{equation}\label{disp-5}
\lambda(q)=0\,.
\end{equation}
From (\ref{disp-3-cc}) it follows that the scalar function $\lambda(q)$ is a 4-th order homogeneous polynomial of the wave covector $q_i$.  It means that, for an arbitrary linear response medium,  there are at most {\it two independent quadratic wave cones} at  every points of the space. This non-trivial physical  fact was previously observed in \cite{birkbook}. 

Observe that in a widely used non-covariant treatment  of the characteristic  system (\ref{disp-1}), see, for instance, \cite{perlick-book}),   one obtains instead of  (\ref{disp-5}) the equation of the form $\omega^2\lambda(q)=0$. In an addition to two light cones mentioned above, this equation has   a  degenerate solution $q_i=(0,k_1,k_2,k_3)$, called {\it zero frequency electromagnetic wave.} In our covariant description, as well as in \cite{birkbook}, this unpleasant solutions are absent.  Thus the problem how to interpret or how to remove the zero frequency waves is not emerges in our approach at all. 

For different explicit forms of the coefficients of the quartic form $\lambda(q)$ called Tamm-Rubilar tensor, see \cite{birkbook}, \cite{Obukhov:2000nw}, \cite{Rubilar:2007qm}, \cite{Itin:2009aa}. 
To our opinion, it is  more convenient to work with the generic form ${\rm adj}(M)=0$ for qualitative analysis and even for explicit calculations. We will demonstrate it in the sequel.  
\subsection{Skewon part of the dispersion relation}
We discuss now how the skewon part contributes to the dispersion relation. 
Substituting   the decomposition of the optic tensor $M^{ij}=P^{ij}+Q^{ij}$ into Eq.(\ref{disp-3-cc}) we have
\begin{eqnarray}\label{disp-8}  
A_{ij}&=&\frac 1{3!}\ep_{ii_1i_2i_3}\ep_{jj_1j_2j_3}\big(P^{i_1j_1}P^{i_2j_2}
P^{i_3j_3} +\nonumber\\ &&
3P^{i_1j_1}Q^{i_2j_2}Q^{i_3j_3}+3P^{i_1j_1}P^{i_2j_2}Q^{i_3j_3} +
\nonumber\\ &&\qquad\qquad\qquad\qquad\,\,\,\,\,
 Q^{i_1j_1}Q^{i_2j_2}Q^{i_3j_3}\big)\,.
\end{eqnarray}
The left hand side of this equation is a symmetric  matrix, thus the antisymmetric matrices  in its right hand side must vanish. 
Indeed, we can check straightforward that the identities 
\begin{eqnarray}\label{disp-10}
&&\ep_{ii_1i_2i_3}\ep_{jj_1j_2j_3}Q^{i_1j_1}Q^{i_2j_2}Q^{i_3j_3}\equiv 0\,,\\
\label{disp-11}
&&\ep_{ii_1i_2i_3}\ep_{jj_1j_2j_3}P^{i_1j_1}P^{i_2j_2}Q^{i_3j_3}\equiv 0\,
\end{eqnarray}
hold for an arbitrary symmetric matrix $P$ and antisymmetric matrix $Q$. 
Note that Eq.(\ref{disp-10}) represents the mentioned fact that the adjoint of an antisymmetric  matrix $Q$ vanishes identically. 
Thus we remain in (\ref{disp-8}) with the relation 
\begin{eqnarray}\label{disp-12}  
\lambda (P,Q) q_iq_j&=&\frac 1{3!}\ep_{ii_1i_2i_3}\ep_{jj_1j_2j_3}\big(P^{i_1j_1}P^{i_2j_2}
P^{i_3j_3} +\nonumber\\ &&\qquad
 3P^{i_1j_1}Q^{i_2j_2}Q^{i_3j_3} \big)=0\,.
\end{eqnarray}
The first term in the right hand side is the adjoint of the symmetric matrix $P^{ij}$. 
Consequently, the dispersion relation takes the form
\begin{equation}\label{disp-12x}  
\lambda(P)q_iq_k +
\frac 1{2!} \ep_{ii_1i_2i_3}\ep_{jj_1j_2j_3}P^{i_1j_1}Q^{i_2j_2}Q^{i_3j_3}=0 \,, \,\,
\end{equation}
where $\lambda(P)$ is the quartic form evaluated on the principle part only. The second term of (\ref{disp-12x}) is calculated in  Appendix. The result is surprisingly simple: 

\vspace{0.2 cm}
 {\bf Proposition 1:} {\it For a most generic linear constitutive pseudo-tensor $\chi^{ijkl}$, the  dispersion relation reads}
\begin{equation}\label{disp-24}  
\boxed{\lambda(P)+P^{ij}Y_iY_j=0\,.}
\end{equation}

Observe some immediate consequences of this formula: 
\begin{itemize}
\item[(1)] The skewon modification of the light cone is provided by a quadratic form $P^{ij}Y_iY_j$.  
\item[(2)] Due to the identity $P^{ij}q_i= 0$, Eq.(\ref{disp-24})   is invariant under a gauge transformation of the skewon covector $Y_i\to Y_i+\theta q_i$.  
\item[(3)] If there is a solution $q_i$ of Eq.(\ref{disp-24}) for which $Y_i(q)\sim q_i$, then $\lambda(P)=0$, i.e., the wave vector $q_i$ lies on the non-modified light cone.
\item[(4)] If $Y_i(q)\sim q_i$ for all solutions of  Eq.(\ref{disp-24}), then the corresponding  skewon  does not modify the light cone structure.
\item[(5)] If $\lambda(P)=0$ for all solutions of  Eq.(\ref{disp-24}), then simultaneously $P^{ij}Y_i(q)Y_j(q)=0$ holds. In this case we remain with a subset of the original (non-skewon) light cone. 
\item[(6)] For a non-trivial solution of the dispersion relation, two  scalar terms $\lambda(P)$ and  $P^{ij}Y_i(q)Y_j(q)$ must be of  opposite signs.
\end{itemize}
In order to analyze the contributions of the skewon part to  wave propagation, we restrict in the next section two  media with a  simplest principal part. 
\section{Skewon contributions to the (pseudo) Riemannian vacuum}
Let the  principal part be  constructed from a generic metric tensor of Euclidean or Minkowski signature. The skewon part, however, let be of the most general form. How, in this case, the skewon modifies the standard light cone structure? An extensive analysis of this problem was given by  Obukhov and Hehl in \cite{Obukhov:2004zz}.  Recently Ni \cite{Ni:2013uwa} provided an analysis of skewon effects on cosmic wave propagation in the approximation of small skewon. In \cite{Itin:2013ica}, \cite{Itin:2014rwa}, some new generic  facts about skewon contribution to Minkowski vacuum are presented. 
 We will give a detailed derivation analysis of these results. 
\subsection{Skewon field on a metric space}
Consider  a manifold endowed with a metric tensor $g^{ij}$. For generality, we will not restrict to the Lorentz signature, even the Euclidean  case does not have a  direct physical application. 
Let the principal part of the constitutive pseudo-tensor be presented in the metric-type form \cite{birkbook}
\begin{equation}\label{skew-1}  
{}^{(1)}\chi^{ijkl}=\sqrt{|g|}\left(g^{ik}g^{jl}-g^{il}g^{jk}\right)\,.
\end{equation}
For simplicity, we assume the units for which the dimension factor in (\ref{skew-1}) is neglected, i.e., the constitutive tensor $\chi^{ijkl}$ is dimensionless.  Recall that  in SI-system the left hand side of (\ref{skew-1}) comes with an additional factor $\lambda_0=\sqrt{\varepsilon_0/\mu_0}$, see  \cite{birkbook} for the details. 

From (\ref{skew-1}),  the principal optic tensor reads
 \begin{equation}\label{skew-2}  
P^{ik}=\sqrt{|g|}\left(g^{ik}q^2-q^iq^k \right)\,.
\end{equation}
Here the indices are raised by the metric $g^{ik}$, i.e., 
\begin{equation}\label{skew-3}  
q^i=g^{ij}q_j\,,\qquad q^2=g^{ij}q_iq_j\,.
\end{equation}
First we  calculate the adjoint (\ref{disp-5}) of the  symmetric matrix  (\ref{skew-2}).  
It takes the form 
 \begin{equation}\label{skew-4}  
\lambda(P)q_iq_j\!=\!\frac 1{3!} \ep_{ii_1i_2i_3} \ep_{jj_1j_2j_3} q^4g^{i_2j_2}g^{i_3j_3}\!\!
\left(q^2g^{i_1j_1}
\!\!-3q^{i_1}q^{j_1}\right).
\end{equation}
Using the standard identities for contraction of the  $\ep$-symbols with the metric tensor, we obtain in (\ref{skew-4})
 \begin{eqnarray}\label{skew-4A}  
\lambda(P)q_iq_j= {\rm sgn}(g)\sqrt{|g|}q^4 q_iq_j\,.
\end{eqnarray}
Thus the principle part  contribution to the dispersion function reads 
\begin{equation}\label{skew-4x}  
\lambda(P) ={\rm sgn}(g)\sqrt{|g|}q^4\,.
\end{equation}
 The contribution of the skewon part follows straightforward from (\ref{skew-1}) 
\begin{equation}\label{skew-12x}  
P^{ij}Y_iY_j=\sqrt{|g|}\left(Y^2q^2-(Y,q)^2\right)\,.
\end{equation}
where the scalar product is defined by the use of the metric, $(q,Y)=g^{ij}Y_iq_j$.  
Substituting into (\ref{disp-24}) we remain with 
\begin{equation}\label{skew-13}
\lambda=\sqrt{|g|}\left({\rm sgn}(g)q^4  +Y^2q^2-(Y,q)^2\right)\,.
\end{equation}
Thus we proved the following 

\vspace{0.2 cm}
 {\bf Proposition 2:} {\it The dispersion relation for  the most generic  skewon  media in the (pseudo) Riemannian  vacuum is represented as}
\begin{equation}\label{skew-14}
\boxed{{\rm sgn}(g)q^4  +Y^2q^2-(Y,q)^2 =0\,.}
\end{equation}

\subsection{Lorenz-type gauge}
The dispersion relation (\ref{skew-14}) may   be given even in a more simpler form. Since the covector $Y_i$ is defined only up to an arbitrary addition of the covector $q_i$, we can  apply an arbitrary scalar gauge condition. 
On a space endowed with a metric  it  can be used in a  form similar to the Lorenz gauge condition
\begin{equation}\label{skew-15}
(Y,q) =g^{ij}Y_iq_j=0\,.
\end{equation}
Note that this condition is applicable for  an arbitrary signature  of the metric tensor, even it is usually used for the Lorentz metric. 
With this expression at hand, the dispersion relation may be used in  the   form  of the following system
\begin{equation}\label{skew-16}
{\rm sgn}(g)q^4 +q^2Y^2=0\qquad {\rm and}\qquad (q,Y)=0\,.
\end{equation}

Let us analyze how the gauge condition (\ref{skew-15}) can be realized on a metric space. 
If a solution $q_i$ of Eq.(\ref{skew-14})  satisfies the relation $q^2=0$, then relation (\ref{skew-15}) is a direct consequence of Eq.(\ref{skew-14}) for an  arbitrary covector $Y$. Let now $q_i$ be a non-light solution of 
Eq.(\ref{skew-14}),  i.e.,  $q^2\ne 0$. 
As in (\ref{mat-0}), we consider a generic expression of the covector $Y$ in terms of the tensor $S$  
\begin{equation}\label{skew-16a}
Y_i=S_i{}^jq_j+\theta q_i\,.
\end{equation}
Recall that $S_i{}^j$ is independent of $q$, while the scalar $\theta$  can be an arbitrary function of $q_i$. Its values  correspond to different choices of the gauge condition.   
Multiplying both sides of (\ref{skew-16a}) by $q$ and requiring (\ref{skew-15}) we obtain 
\begin{equation}\label{skew-16b}
(Y,q)=S^{ij}q_iq_j+\theta q^2=0\,.
\end{equation}
Consequently, for $q^2\ne 0$, 
\begin{equation}\label{skew-16b-cc}
\theta=-\frac{S^{mn}q_mq_n}{q^2}\,.
\end{equation}
Thus for $q^2\ne 0$,  the optic covector is expressed in the Lorenz-type gauge as 
\begin{equation}\label{skew-16c}
Y_i=S_i{}^jq_j-\frac{S^{mn}q_mq_n}{q^2}q_i\,.
\end{equation}
Consequently, the Lorenz-type condition $(Y,q)=0$ may be applied for an arbitrary tensor $S_i{}^j$ and for an arbitrary wave covector $q_i$.

\subsection{Euclidean signature}
For a positive signature of the metric,  the dispersion relation in (\ref{skew-16}) takes the form
\begin{equation}\label{skew-Eu}
q^4  +Y^2q^2 =0\,.
\end{equation}
Both terms in (\ref{skew-Eu}) are nonnegative. Consequently any solution of (\ref{skew-Eu}) must satisfy $q^2=0$.  For the Euclidean metric, this  yields the unique trivial solution, $q_i=0$. 
The gauge condition $(q,Y)=0$ also holds. 
Thus we come to the following  conclusion:

\vspace{0.2 cm}
 {\bf Proposition 3:} 
{\it For a Euclidean signature metric space endowed with an arbitrary skewon, the dispersion relation has the unique trivial solution, $q_i=0$.}

\vspace{0.2 cm}
 Thus an arbitrary skewon part cannot modify the elliptic character of the Euclidean metric. Wave propagation was forbidden in an Euclidean signature model and it remains forbidden even when an arbitrary skewon field is "switched on".

\subsection{Lorentz  signature}
Also for a negative signature of the metric,  it is more convenient to use the dispersion relation in the Lorenz gauge. Thus we are dealing with the system 
\begin{equation}\label{skew-Lo}
q^4  =Y^2q^2\,\quad {\rm and}\quad (Y,q)=0\,.
\end{equation}
In order to analyze its solutions, we first observe that the left hand side of the first equation is nonnegative, $q^4\ge 0$. If $q^2\ne 0$, then it follows that $Y^2q^2>0$. Thus the factors  $q^2$ and $Y^2$  have  the same sign. This means that the covectors $Y_i$ and $q_i$ must be both timelike or both spacelike. But two timelike covectors cannot be orthonormal, as  required by the second equation of (\ref{skew-Lo}).  Consequently, if $q_i\ne 0$, the covectors  $Y_i$ and $q_i$ must be  both spacelike.  
So we proved the following

\vspace{0.2 cm}
{\bf Proposition 4:}  
{\it In the Lorentz signature metric space, the solution of the skewon modified dispersion relation  can be spacelike or null,
\begin{equation}\label{res}
q^2\le 0\,.
\end{equation}
In other words, an arbitrary skewon widens the light cone. Thus the  skewon contribution to a generic pseudo-Riemannian vacuum   media increases the light velocity. }

\section{Antisymmetric skewon}
Up to now our qualitative consideration did not assume any   specific form of  skewon. We will now consider  some explicit examples of electromagnetic wave propagation in skewon media.  For certain special cases, a detailed analysis  was provided in  \cite{Obukhov:2004zz}. We will study how these  results are embedded in our formalism and how they can be extended. We will also discuss the results of the approximated analysis provided recently by Ni, \cite{Ni:2013uwa}.

A simplest way to construct a skewon is to start with an arbitrary  traceless  tensor $S_i{}^j$ which is completely equivalent to  $^{(2)}\chi^{ijkl}$.  On a  space endowed with a metric tensor $g_{ij}$, the mixed   tensor  $S_i{}^j$ can be replaced by the covariant and contravariant tensors
\begin{equation}\label{es-0}
S^{ij}=g_{ik}S_k{}^j\,,\quad{\rm and}\quad 
S_{ij}=g_{jk}S_i{}^k \,.
\end{equation}
Note that these two  tensors depend not only of the skewon itself but also of the metric. One advantage of   the tensors (\ref{es-0}) is that they  can be invariantly decomposed into   symmetric and skew-symmetric parts. Since these two parts do not mix under coordinate transformations,  they can be studied separately. 

In this section we consider  the case of a  pure antisymmetric skewon tensor
\begin{equation}\label{es-1}
S_{ij}=-S_{ji}\,.
\end{equation}

\subsection{Dispersion relation}
We observe first that the antisymmetric skewon is, in fact, a simplest type of a skewon. 
Indeed, the traceless condition for such tensor holds identically, 
\begin{equation}\label{es-2}
S_{ij}g^{ij}=0\,.
\end{equation}
Moreover,
\begin{equation}\label{es-3}
S_{ij}q^iq^j=0\,.
\end{equation}
Then  when the antisymmetric tensor is substituted into Eq.(\ref{skew-16c}), the non-regular term  is canceled.  
 Consequently we can realize the Lorentz gauge condition $(Y,q)=0$ with  a specially simple skewon covector $Y$ that is linear in the wave covector $q$
\begin{equation}\label{es-4}
Y_i=S_{ij}q^j\,.
\end{equation}
We observe that the Lorentz gauge condition is satisfied automatically. 
In the dispersion relation
\begin{equation}\label{es-5}
q^4 =q^2Y^2\,
\end{equation}
 $Y(q)$ is a polynomial now. Hence this equation is  decomposed into two independent equations 
\begin{equation}\label{es-6a}
q^2=0\,,\quad {\rm or }\quad
q^2 =Y^2\,.
\end{equation}
The later equation is given in an implicit form (recall that $Y$ is linear in $q$). Since $Y_i$ is a first order homogeneous (even linear) function of $q_i$, this equation represents an algebraic cone.

The fact of splitting of the dispersion relation for the antisymmetric skewon   was derived   in \cite{Obukhov:2004zz}.  There, the decomposition of the  dispersion relation into two independent equations  was interpreted as the {\textit {birefringence effect}} known from crystal optics. 
This statement requires however a slightly more precise analysis.  
In particular, it is not clear if the equation $q^2 =Y^2$ has real solutions and if these solutions  represent  an {\textit { observable light cone. }} We will discuss these questions in the following section. 

\subsection{Light-cone  condition}
In order to  represent a light cone, Eq.(\ref{es-6a}) must  have {\it non-zero  real solutions.} We will prove now that this is indeed the case for an  almost  arbitrary antisymmetric skewon.

Substituting (\ref{es-4})   the equation $q^2-Y^2=0$  we derive
\begin{equation}\label{es-10}
\left(g_{ij}-S_{im}S_{jn}g^{mn}\right)q^iq^j=0\,.
\end{equation}
It can be expressed in terms of an effective metric  ${\hat g}_{ij}$
\begin{equation}\label{es-8}
{\hat g}_{ij}q^iq^j=0\,,
\end{equation}
where 
\begin{equation}\label{es-9}
{\hat g}_{ij}=g_{ij}-S_{im}S_{jn}g^{mn}\,.
\end{equation}

On a four dimensional manifold, the metric ${\hat g}_{ij}$  can be: 
\begin{itemize}
\item[(i)] Euclidean with the signature $(+,+,+,+)$, 
\item[(ii)] Lorentzian  with the signature $(+,-,-,-)$, 
\item[(iii)] of  mixed type  with the signature $(+,+,-,-)$, 
\item[(iv)] degenerate with ${\rm{det\,}}  g=0$. 
\end{itemize}
Only the Lorentz case provides  a  light cone. Observe that on the 4-dimensional manifold,  the metric has Lorentz signature if and only if its determinant is negative. Consequently we have a {\textit{necessary condition for birefringence }}
\begin{equation}\label{es-11}
{\rm det}\left(g_{ij}-S_{im}S_{jn}g^{mn}\right)<0\,.
\end{equation}
This relation can be modified being multiplied by ${\rm det}g^{ik}$. Since $g$ is a Lorentz signature metric,  we obtain an equivalent condition 
\begin{equation}\label{es-12}
{\rm det}\left(\d^i_j+S^{i}{}_mS^m{}_j\right)>0\,.
\end{equation}
Note that in this form, the condition is independent of the metric tensor and applicable in an even  more generic case.  

Since the inequality (\ref{es-11}) is pure algebraic, i.e., point-wise, we can apply coordinate transformations to replace the metric $g_{ij}$ by the flat Minkowski metric  $g_{ij}={\rm diag}(1,-1,-1,-1)$. 
We  use the standard "electromagnetic parametrization' of the antisymmetric matrix 
\begin{equation}\label{es-13}
S_{01}=\a_1\,, \qquad S_{02}=\a_2\,, \qquad S_{03}=\a_3\,,
 \end{equation}
and 
\begin{equation}\label{es-14}
S_{23}=\b_1\,, \qquad S_{13}=-\b_2\,, \qquad S_{12}=\b_3\,.
 \end{equation}
 With this parametrization, the effective metric takes the form
\begin{widetext}
\begin{equation}\label{es-15}
{\hat g}_{ij}=\left( \begin{array}{cccc}
1+\a_1^2+\a_2^2+\a_3^2 & \a_2\b_3-\a_3\b_2 & \a_3\b_1-\a_1\b_3 &\a_1\b_2-\a_2\b_1 \\
\a_2\b_3-\a_3\b_2 &-1+\b_2^2 +\b_3^2-\a_1^2&-\a_1\a_2-\b_1\b_2& -\a_1\a_3-\b_1\b_3\\
\a_3\b_1-\a_1\b_3 & -\a_1\a_2-\b_1\b_2 & -1+\b_1^2 +\b_3^2-\a_2^2&-\a_2\a_3-\b_2\b_3\\
\a_1\b_2-\a_2\b_1&-\a_1\a_3-\b_1\b_3 &-\a_2\a_3-\b_2\b_3 &-1+\b_1^2 +\b_2^2-\a_3^2
\end{array} \right)
 \end{equation}
\end{widetext}
We calculate the determinant of this matrix with the the computer
algebra package Reduce-Excalc, see
\cite{Hearn}. The result is surprisingly simple 
\begin{equation}\label{es-16}
{\rm det}({\hat g}_{ij}) =-\left[(\a\cdot\b)^2+(\b^2-\a^2-1)\right]^2\,.
 \end{equation}
Here the standard scalar product notations are used, $(\a\cdot\b)=\a_1\b_1+\a_2\b_2+\a_3\b_3$, and  
$\a^2=\a_1^2+\a_2^2+\a_3^2$, and similarly for $\b^2$. 
 The following  statement results from Eq.(\ref{es-16}).  

\vspace{0.1 cm}

{\bf Proposition 5:} {\it For an arbitrary antisymmetric skewon, the determinant of the effective metric is non-positive.  Thus, the equation $q^2=Y^2$ has non-zero real solutions. In the case $(\a\cdot\b)^2\ne \a^2-\b^2+1$, the effective metric has Lorentz signature.}

As a side result  we derived a degenerate skewon with the parameters  $(\a\cdot\b)^2= \a^2-\b^2+1$. We will study it in the sequence. 
\subsection{The form of the modified light cone}
The solution $q^2=0$ represents a perfect light cone. Using the standard notation $q_i=(\omega,{\rm k}_1,{\rm k}_2,{\rm k}_3)$ we observe its basic  properties: 
\begin{itemize}
\item[(1)] The unique vertex is localized at the origin $q_i=0$;
\item[(2)] The symmetry axis is directed along the $\omega$-axis;
\item[(3)]  The timelike sections $\omega={\rm const}$ is a sphere with   full $SO(3)$ symmetry.
\end{itemize}
What can be said in this context about the second cone 
$q^2=Y^2?$

 The origin point $q_i=0$ is evidently a solution of this equation, because  $Y_i$ is a 1-st order homogeneous polynomial in $q_i$. This fact also yields  the algebraic cone structure: For every solution $q_i$, the scaled covector $Cq_i$ is also a solution. The only point that remains fixed under this rescaling transformation is the origin $q_i=0$. Thus the hypersurface is an algebraic cone with a unique vertex localized at the origin.

In order to understand the position of the cone axis we need the space-time decomposed form of  the equation $q^2=Y^2$. Using the effective metric (\ref{es-15}) we derive
\begin{equation}\label{mod-cone1}
(1+\a^2)\omega^2+2\omega\left({\rm k}\cdot [\a\times\b]\right)-{\rm k}^2-(\a\cdot {\rm k})^2+(\b\times {\rm k})^2=0\,,
\end{equation}
where the vector (cross) and scalar (dot) products of 3-vectors are used. 
We restrict  our analysis of this algebraic equation to two special cases.

For $\b=0$, we are left in (\ref{mod-cone1}) with
\begin{equation}\label{mod-cone2}
(1+\a^2)\omega^2={\rm k}^2+(\a\cdot {\rm k})^2\,.
\end{equation}
We observe that the right hand side of this equation is non-negative. Consequently,  for constant values of $\omega$,  (\ref{mod-cone2}) represents a compact second order polynomial hypersurface. Thus it is an ellipsoid. 
Moreover, if $(\omega, k)$ is a solution than $(\omega, -k)$ is also a solution. Thus the line $k=0$ is a symmetry axis of the cone. 

For $\a=0$, Eq.(\ref{mod-cone1}) takes the form 
\begin{equation}\label{mod-cone3}
\omega^2-{\rm k}^2+(\b\times {\rm k})^2=0\,.
\end{equation}
There are  two possibilities:

If $||b||\le 1$, we rewrite Eq.(\ref{mod-cone3}) as
\begin{equation}\label{mod-cone4}
\omega^2=(1-\b^2){\rm k}^2+(\b\cdot{\rm k})^2\,.
\end{equation}
The right hand side here is non-negative. In this case, all sections of the surface with the constant $\omega$ are ellipsoids. Here the symmetry axis of the cone  is the line $k=0$ as usual. 

If $||b||> 1$, we rewrite Eq. (\ref{mod-cone3}) as 
\begin{equation}\label{mod-cone5}
\omega^2+(\b^2-1){\rm k}^2=(\b\cdot{\rm k})^2\,.
\end{equation}
Now the left hand side is non-negative. 
Thus the sections with the constant values of $(\b\cdot{\rm k})$ are ellipsoids. 
Moreover, the symmetry axis of the 3 dimensional cone is lying in hyperplane $(\b\cdot{\rm k})$, i.e., it is normal to the $\omega$-axis. 

In other words, we identify  a new feature of this model: The time and spatial axes are interchanged when the parameter $\b^2$ crosses the value 1. 
It has some  similarity to the Schwarzschild solution in ordinal coordinates when  the   time and radial axes interchange at the critical radius. 
\subsection{Degenerate skewon}
Here we consider some special sets of skewon parameters which disregarded and which lead to  degenerate cases. 
The effective metric is singular when 
\begin{equation}\label{es-18a}
{\rm det}\left(\d^i_j+S^{im}S_{mj}\right)= 0\,.
 \end{equation}
In  parametric form, this equation becomes 
\begin{equation}\label{es-18}
(\a\b)^2+\b^2-\a^2-1=0\,.
 \end{equation}
Due to the Cauchy-Schwarz inequality, $ (\a\b)^2 \le\a^2\b^2$, this equation has real solution only if 
\begin{equation}\label{es-19}
\b^2\ge 1\,.
 \end{equation}
We consider, for instance, a special solution of Eq.(\ref{es-18}) of the form 
\begin{equation}\label{es-20}
\a=0\,, \qquad \b^2= 1\,.
 \end{equation}
In this case, the effective metric (\ref{es-15}) reads 
\begin{equation}\label{es-21}
{\hat g}_{ij}=\left( \begin{array}{cccc}
1& 0 & 0 &0 \\
0 &-\b_1^2&-\b_1\b_2& -\b_1\b_3\\
0 & -\b_1\b_2 & -\b_2^2&-\b_2\b_3\\
0 &-\b_1\b_3 &-\b_2\b_3 &-\b_3^2
\end{array} \right)\,.
 \end{equation}
Consequently, the dispersion relation is given by
\begin{equation}\label{es-22}
\omega^2-(\b_1{\rm k}_1+\b_2{\rm k}_2+\b_3{\rm k}_3)^2=0\,,
 \end{equation}
or, equivalently, by the two linear equations
\begin{equation}\label{es-22x}
w=\pm(\b_1{\rm k}_1+\b_2{\rm k}_2+\b_3{\rm k}_3)\,.
 \end{equation}
This linear dispersion relation must  not be confused with the ordinary linear dispersion in vacuum, which is of the form $w=k=\sqrt{{\rm k}_1^2+ {\rm k}_2^2+{\rm k}_3^2}$. 
In Eq.(\ref{es-22x}),  we are left  with two hyperplanes  instead of a hypercone. The effective velocity of the wave is not bounded on such a "wavefront." It means that this special case provides a very hard violation  of the special relativity principles.

\subsection{Optic axes generated by an antisymmetric  skewon}
When the degenerate case is excluded, we are left with two perfect algebraic cones with the same vertex at the origin.  The mutual arrangement of these cones can  however be of various topological types. Let us first study whether these algebraic cones are intersect and what is the  set of their intersection. 

We are looking for a  non-zero solution of the system  
\begin{equation}\label{es-ax18}
q^2=0\,,\qquad{\rm and}\qquad Y^2=q^2\,.
 \end{equation}
 Due to homogeneity, every solution of Eq.(\ref{es-ax18}) describes a 2-dimensional plan in 4-dimensional space-time. The  spatial  projection of this plan is  a  direction in which a wave propagates without birefringence, i.e., the {\it optic axis}. The system (\ref{es-18}) is equivalent to 
\begin{equation}\label{es-ax19}
q^2=0\,,\qquad{\rm and}\qquad Y^2=0\,.
 \end{equation}
With these simple equations at hand, we can prove the following:

\vspace{0.2 cm}

{\bf Proposition 6:} {\it In a pseudo-Riemannian manifold endowed with an arbitrary   antisymmetric skewon, the wavefront has two separate optic axes. }

\vspace{0.2 cm}

{\bf Proof: }
In order to have a non-trivial solution of system (\ref{es-ax18}), both covectors $Y_i$ and $q_i$ must be   light-like relative to the basic metric $g_{ij}$. We recall  that these two covectors also must  satisfy the gauge condition  
\begin{equation}\label{es-ax20}
(qY)=0\,,
 \end{equation}
i.e., they must be pseudo-orthonormal. It is well known, see e.g. \cite{Giulini:2006uy},  that two light-like vectors are pseudo-orthonormal if and only if they are proportional. Thus the optic axis covector $q$ must satisfy the system
\begin{equation}\label{es-ax21}
q^2=0\,,\qquad{\rm and}\qquad Y_i(q)=mq_i\,
 \end{equation}
with some real parameter $m$. 
Let us derive  the values of the parameter $m$. In terms of the skewon tensor $S_{ij}$, the  equation $ Y=mq $ takes the form of an eigenvector  problem
\begin{equation}\label{es-ax22}
(S_{ij}-mg_{ij})q^j=0\,.
 \end{equation}
This system has a non-trivial solution if and only if 
\begin{equation}\label{es-ax23}
{\rm det}(S_{ij}-mg_{ij})=0\,.
 \end{equation}
Every real eigenvalue $m$  produces a real eigenvector $q_i$ of Eq.(\ref{es-ax22}). Moreover, distinct real eigenvalues  correspond to linear independent real eigenvectors. 
We observe also that,  due to the gauge relation $(q,Y)=0$, every covector corresponded to a non-zero eigenvalue $m$ is necessary light-like. Indeed,  the equation $Y=mq$ results into $mq^2=0$. Note that, in general, this is not correct  for $m=0$. Thus we must give a special treatment of  the case of a zero eigenvalue.

We shall now derive  the solutions of Eq.(\ref{es-ax23}).
Using the "electromagnetic parametrization" (\ref{es-13}) we have
\begin{equation}\label{es-ax24}
S_{ij}-mg_{ij}= \left( \begin{array}{cccc}
m &\a_1 & \a_2&\a_3 \\
-\a_1 & -m & \b_3&-\b_2\\
-\a_2 & -\b_3 & -m &\b_1\\
-\a_3&\b_2&-\b_1&-m
\end{array} \right)\,.
 \end{equation}
Calculating the determinant of this matrix we obtain the characteristic equation 
\begin{equation}\label{es-ax25}
{\rm det}\left(S_{ij}-mg_{ij}\right)= m^4+m^2(\b^2-\a^2)-(\a\b)^4=0\,.
 \end{equation}
 The discriminant of this biquadratic equation  
\begin{equation}\label{es-ax26}
 (\a^2-\b^2)^2+4(\a\b)^4\ge 0\,
 \end{equation}
is non-negative for arbitrary non-zero values of the parameters $\a,\b$. 
Also we observe  that the free term in Eq.(\ref{es-ax25}) is non-positive,  $-(\a\b)^4\le 0$. 
It means:

1) In the case $(\a\b)\ne 0$,  Eq.(\ref{es-ax25}) has  two real  and two purely imaginary  conjugated  solutions. The real solutions are non-zero and expressed as 
\begin{equation}\label{es-ax26a}
m=\pm\left(\frac{\sqrt{(\a^2-\b^2)^2+4(\a\b)^4}
+\a^2-\b^2}2\right)^{1/2}\,. 
 \end{equation}
With these two  real distinct values of the parameter $m\ne 0$, we have two optic axes. 

2) In the case 
\begin{equation}\label{es-ax26cx}
(\a\b)=0\,,\qquad \a^2>\b^2\,,
 \end{equation}
 Eq.(\ref{es-ax25}) has four real solutions -- two non-zero solutions
\begin{equation}\label{es-ax26c}
m=\pm\sqrt{\a^2-\b^2}\,,
 \end{equation}
and a zero solution $m=0$ of multiplicity two. 
As we already observed, every non-zero real eigenvalue $m$ corresponds to  an optic axis. The zero eigenvalue  $m=0$ does not correspond to any optic axis. 
It can be checked by elementary explicit calculations that the conjunction of (\ref{es-ax26cx}) and  (\ref{es-ax20})  only has the  trivial solution. 

So we are left with two optic axes corresponding to the eigenvalues given in Eq.(\ref{es-ax26c}).

3) We consider  now the opposite case: 
\begin{equation}\label{es-ax26cxx}
(\a\b)=0\,,\qquad \a^2<\b^2\,.
 \end{equation}
 The two eigenvalues (\ref{es-ax26c}) are purely imaginary now, and we are left with  $m=0$ of  multiplicity two. Comparing with the previous case, we get now two real  light-like eigenvectors   corresponding
to the eigenvalue $m=0$. Namely, it is enough to chose $q=(\b,0,\sqrt{\b^2-\a^2},\pm\a)$. Once more we have two optic axes. 

4) In the last case,
\begin{equation}\label{es-ax26cxy}
(\a\b)=0\,,\qquad \a^2=\b^2\,,
 \end{equation}
 there is only one eigenvalue $m=0$  of  multiplicity four. It corresponds to  two   eigenvectors  $q=(\b,0,0,\pm\a)$, and consequently there are again two optic axes. 

Thus our proposition is proved. $\blacksquare$

\subsection{Illustrative examples of antisymmetric skewon}
For almost arbitrary parameters of an antisymmetric skewon we derived the existence of two wavefront surfaces, which are tangential one to another along two optic axes. The mutual dispositions of these surfaces may be of topologically different type. In order to identify these topological types, we proceed with some explicit examples.
\subsubsection{Electric type skewon}
Consider a skewon of  pure "electric type" 
 \begin{equation}\label{illustr-ant1}
\a=(\a_1,\a_2,\a_3)\,,\qquad {\rm and}\qquad \b=(0,0,0)\,.
\end{equation}
Here the effective metric (\ref{es-15}) related to the  extraordinary light cone takes the form 
\begin{equation}\label{illustr-ant2}
{\hat g}_{ij}=\left( \begin{array}{cccc}
1+\a^2& 0 & 0 &0 \\
0 &-1-\a_1^2&-\a_1\a_2& -\a_1\a_3\\
0& -\a_1\a_2& -1-\a_2^2&-\a_2\a_3\\
0&-\a_1\a_3 &-\a_2\a_3&-1-\a_3^2
\end{array} \right).
 \end{equation}
The determinant of this metric,  
\begin{equation}\label{illustr-ant2x}
{\rm det}({\hat g}_{ij}) =-\left(\a^2+1\right)^2\,,
 \end{equation}
 is strictly negative  for all values of the parameters $\a$. 

The dispersion relation for the extraordinary wave, ${\hat g}_{ij}q^iq^j=0$, is 
\begin{equation}\label{illustr-ant4}
(1+\a^2)\,\omega^2={\rm k}^2+(\a\cdot {\rm k})^2\,.
 \end{equation}
Since all terms in Eq.(\ref{illustr-ant4}) are positive, all sections of this 3-dimensional cone with  constant values of $\omega$ are 2-dimensional ellipsoids. 

The optic covector reads
\begin{equation}\label{illustr-ant4x}
Y_i=-\big((\a\cdot {\rm k}),\a_1\omega,\a_2\omega,\a_3\omega\big)\,.
 \end{equation}

The two optic axes are straightforward calculated now from (\ref{es-ax21}) with $m=\pm||\a||=\pm\sqrt{\a_1^2+\a_2^2+\a_3^2}$. We have 
\begin{equation}\label{illustr-ant5}
q_i=(\pm||\a||,\a_1,\a_2,\a_3)\,.  \end{equation}
In Figs.1 and 2, we present different sections of two 3-dimensional light cones.

Consequently,  the antisymmetric skewon of  pure "electric type" provides a model  of birefringent    medium.
\begin{figure}[h!]
\parbox [t ]{0.2\textwidth }{
\includegraphics[width=3.5cm]{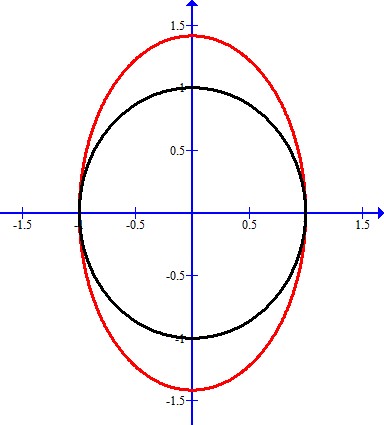}}\qquad
\parbox [t ]{0.2\textwidth }{
\includegraphics[width=3.5cm]{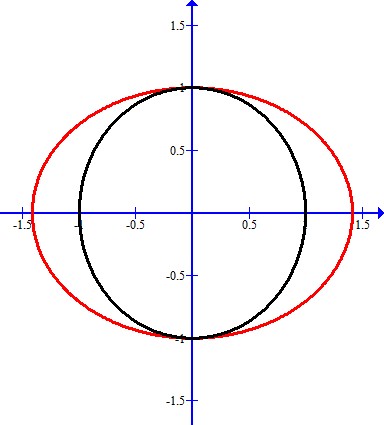}}\qquad
\parbox [t ]{0.20\textwidth }{
\includegraphics[width=3.5cm]{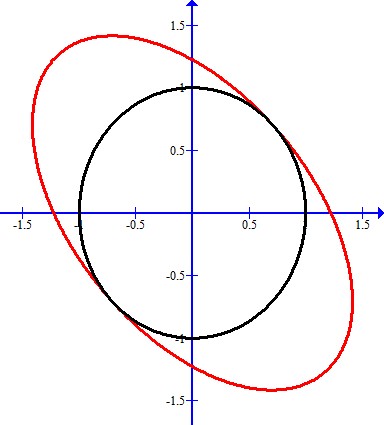}}\qquad 
\hspace{0.01 cm}
\parbox [t ]{0.20\textwidth }{
\includegraphics[width=3.5cm]{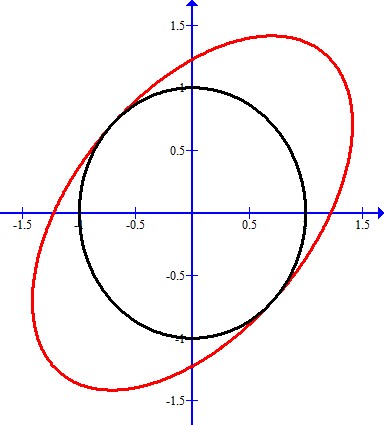}}
\caption[]{Sections of the light cones generated by electric-type skewon. In all cases, $w=1$ and $k_3=0$. Light cones correspond to $\a_i=(1,0,0)$, $\a_i=(0,1,0)$, $\a_i=(1,1,0)$, and $\a_i=(1,-1,0)$, respectively.}
\end{figure}%
\begin{figure}[h!]%


\parbox [t ]{0.5\textwidth }
{
\!\!\!\!\!\!\!\!\!\! \includegraphics[width=8cm]
{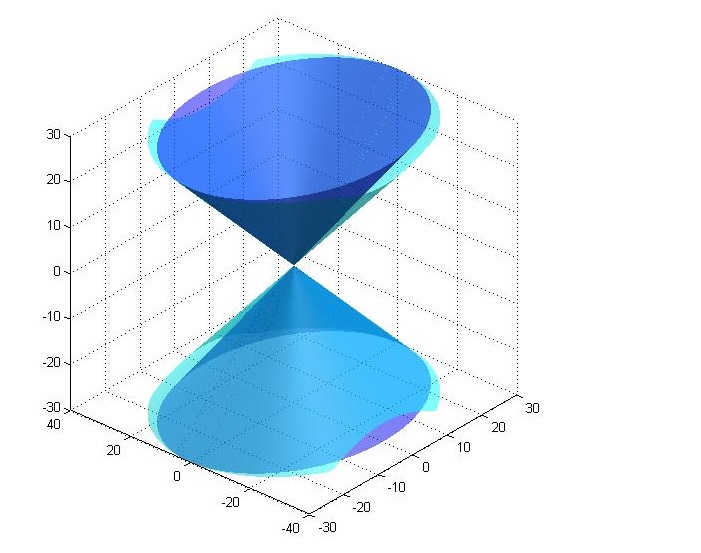}}\qquad

\vspace{0.2cm}

\parbox [t ]{0.40\textwidth }{
\includegraphics[width=7.5cm]{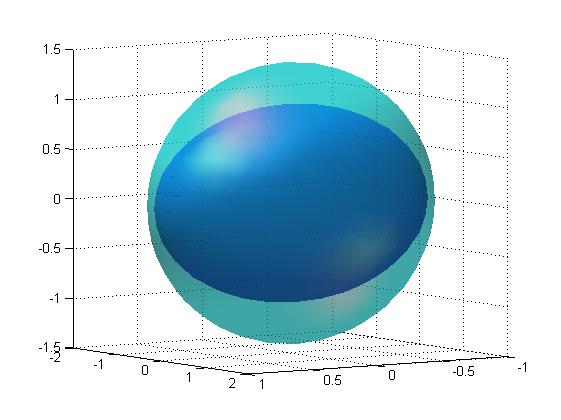}} 
\caption[]{Sections of the light cones generated by an electric-type skewon.  Space images  of the ordinary and extraordinary light cones  and wavefront ellipsoids are generated by the skewon with $\a=(1,0,0)$.  Two axes are represented by the two lines of intersection on the first graph and by the two points  of intersection  on the second one.}
\end{figure}
\subsubsection{Magnetic-type skewon}
We consider now a skewon of  a pure "magnetic type" parametrized by  
 \begin{equation}\label{illustr-ant6}
\a=(0,0,0)\,,\qquad {\rm and}\qquad \b=(\b_1,\b_2,\b_3)\,.
\end{equation}
In this case, the effective metrics (\ref{es-15}) reads
\begin{equation}\label{illustr-ant7}
{\hat g}_{ij}=\left( \begin{array}{cccc}
1 & 0& 0 &0 \\
0 &-1+\b_2^2 +\b_3^2 & -\b_1\b_2& -\b_1\b_3\\
0&-\b_1\b_2 & -1+\b_1^2+\b_3^2&-\b_2\b_3\\
0&-\b_1\b_3 &-\b_2\b_3 &-1+\b_1^2 +\b_2^2
\end{array} \right)
 \end{equation}
The determinant of this metric is
\begin{equation}\label{es-16-cc}
{\rm det}({\hat g}_{ij}) =-\left(\b^2-1\right)^2\,,
 \end{equation}
where $\b^2=\b_1^2+\b_2^2+\b_3^2$. 
The determinant is non-positive for all values of the parameter $\b$ and vanishes only for $|\b|=1$. Consequently, for $|\b|\ne 1$, Eq.(\ref{es-16-cc}) represents a hypercone.  
For $\b^2=1$  the  hypercone degenerates into two hyperplanes. 

Here the dispersion relation of the extraordinary wave is
\begin{equation}\label{illustr-ant8}
\omega^2=(1-\b^2){\rm k}^2+(\b\cdot{\rm k})^2\,.
\end{equation}
From this equation, we conclude:
\begin{itemize}
\item[{(1)}] For $|\b|<1$, the right hand side of this equation is positive, thus a section with  constant value of $w$ is a 2-dimensional  ellipsoid. This is similar to the  ordinary case of birefringence. The extraordinary wave has a velocity greater a speed of light and depends on the spatial direction. 
\item[{(2)}] For   $|\b|=1$, Eq.(\ref{illustr-ant8}) takes the form $w=\pm(\b\cdot {\rm k})$. This is a pair of hyperplans. This is the degenerated case with a non-limiting velocity of the extraordinary wave. 
\item[{(3)}] For   $|\b|>1$, Eq.(\ref{illustr-ant8}) returns to be a hypercone. Its central axis, however, lies in a spatial direction, instead of the time direction. 
\end{itemize}
In all cases, there are two optic axes  given by the equations
\begin{equation}\label{illustr-ant8x}
q=(\pm||\b||,\b_1,\b_2,\b_3)\,,
\end{equation}
where $||\b||=\sqrt{\b_1^2+\b_2^2+\b_3^2}$. 
\begin{figure}[h!]
\parbox [t ]{0.15\textwidth }{
\includegraphics[width=3.5cm]{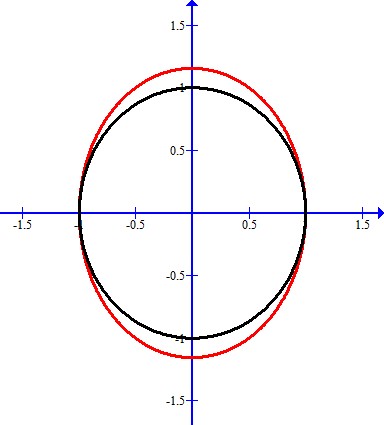}}\qquad
\parbox [t ]{0.15\textwidth }{
\includegraphics[width=3.5cm]{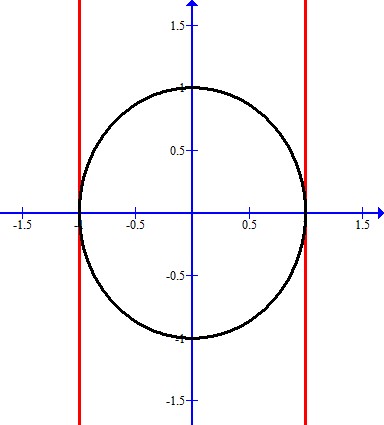}}\qquad
\parbox [t ]{0.15\textwidth }{
\includegraphics[width=3.5cm]{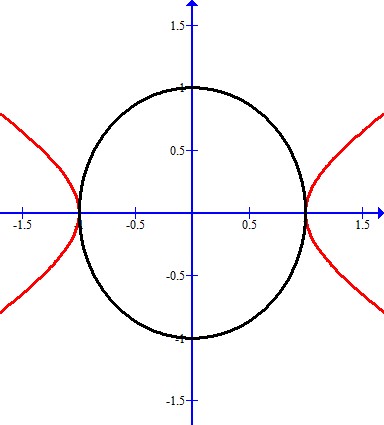}}\qquad
\parbox [t ]{0.15\textwidth }{
\includegraphics[width=3.5cm]{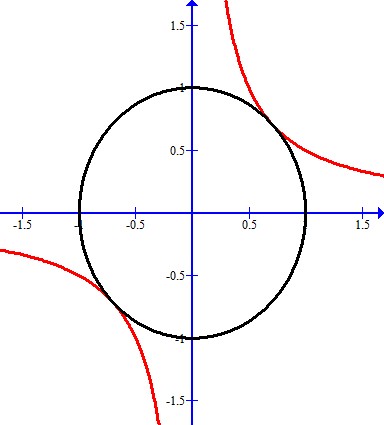}}
\caption[]{Sections of the light cones generated by electric-type skewon. In all cases, $w=1$ and $k_3=0$. Light cones correspond to $\a_i=(1,0,0)$, $\a_i=(0,1,0)$, $\a_i=(1,1,0)$, and $\a_i=(1,-1,0)$, respectively. The optic axes are represented by the lines of intersection of the cones.}

\end{figure}
\begin{figure}[h!]
\parbox [t ]{0.22\textwidth }{
\includegraphics[width=4.5cm]
{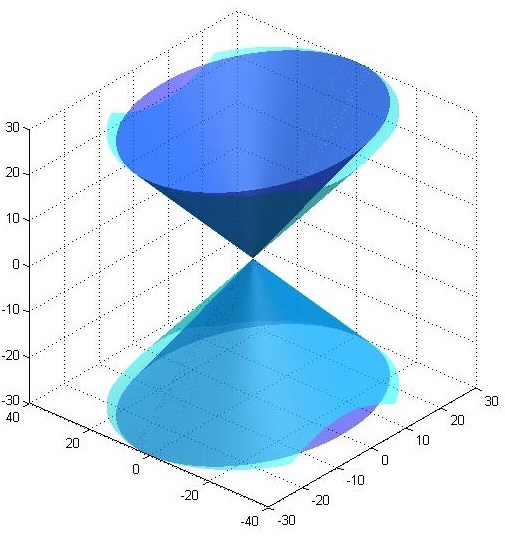}
}
\parbox [t ]{0.22\textwidth }{
\includegraphics[width=5cm]{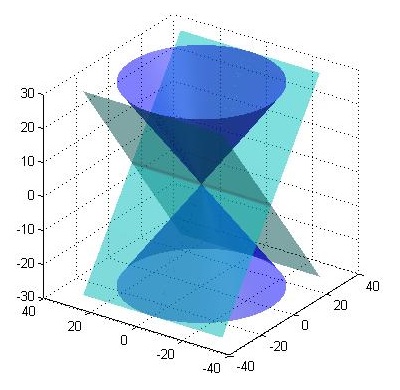}}
\qquad
\parbox [t ]{0.50\textwidth }{\qquad
\includegraphics[width=6cm]{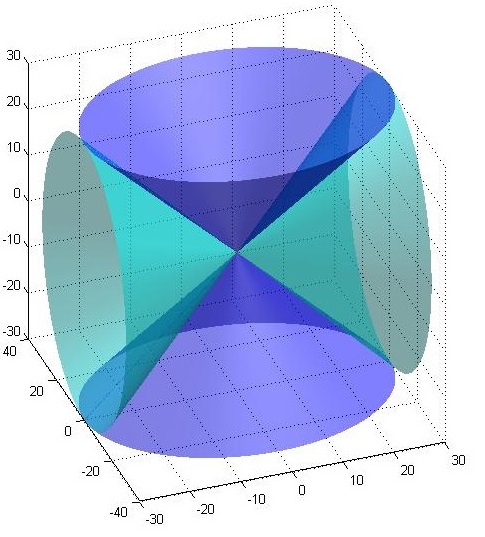}}
\caption[]{Sections of the light cones generated by the magnetic-type skewon.  Space images  with ${\rm k}_3=0$ are generated by  the skewons  $\b_i=(1/2,0,0)$, $\b_i=(1,0,0)$, $\b_i=(2,0,0)$, respectively. Two optic axes are represented by the two lines of intersection. 
 }
\end{figure}

Consequently, only for small values of the magnetic type  skewon parameters $\b<1$ the ordinary type of birefringence emerges. Higher values of $\b$ yields a rather pathological violation of Lorentz symmetry. 
\section{Symmetric skewon}
When skewon is considered on a space endowed with a metric tensor, a special symmetric traceless skewon can be defined.  In terms of the tensor $S_{ij}$, it satisfies the relations $S_{ij}=S_{ji}$ and $S^i{}_i=0$. In a 4-dimensional space, such a tensor  has 9 independent components. 
\subsection{Parametric gap}
Consider the dispersion relation for a generic skewon in the form
\begin{equation}\label{par-1}
q^4-q^2Y^2+(q,Y)^2=0\,.
\end{equation}
When the gauge condition $\theta=0$ is applied in (\ref{skew-16a}), every term in Eq.(\ref{par-1}) is polynomial in the wave covector $q$. Consequently we can consider a  solution of Eq.(\ref{par-1}) in the form 
\begin{equation}\label{par-2}
q^2=\frac 12 \left(Y^2\pm \sqrt{Y^4-(q,Y)^2}\right)\,.
\end{equation}
The Lorentz square $q^2$ must be real, consequently the expression under the square root must be non-negative. Recall that in the case of an antisymmetric matrix $S_{ij}$, the term  $(q,Y)$ was equal zero also in for the gauge $\theta=0$. Thus the term under the square root was non-negative, $Y^4\ge 0$,  and there were two real solutions for the variable $q^2$. In the case of a symmetric skewon, the situation changes crucially. Now both terms  under the square root are non-zero and the inequality 
\begin{equation}\label{par-3}
Y^4-(q,Y)^2\ge 0\,
\end{equation}
must hold. Recall that the covector $Y$ is linear in the skewon matrix $S_{ij}$. If we rescale this matrix, $S_{ij}\to CS_{ij}$, Eq.(\ref{par-3}) takes the form
\begin{equation}\label{par-4}
C^2Y^4-(q,Y)^2\ge 0\,.
\end{equation}
We observe that for  small values of the parameter $C$, the first term approaches  zero and the inequality is broken. In fact, the same is true even for a generic skewon since 
the term $(q,Y)$ is proportional only to the symmetric part of the skewon.
Consequently, we proved, \cite{Itin:2014rwa}

\vspace{0.2 cm}
 {\bf Proposition 7: }{\it  Let the skewon have a non-zero symmetric part. For sufficient small nonzero magnitudes of the skewon parameters  the dispersion relation has no solutions. }

\subsection{Vector parametrization}
It is convenient to represent the tensor $S_{ij}$ in term of  vector fields \cite{Obukhov:2004zz}. We call such a representation a  {\it vector parametrization}. Let us consider a vector field $v_i$. We can write a symmetric traceless  tensor in the form
\begin{equation}\label{sym-1}
S_{ij}=v_iv_j-\frac 14 v^2g_{ij}\,.
\end{equation}
Here the square of the vector $v$ is defined by  the metric tensor,  $v^2=g^{ij}v_iv_j$. 
Certainly one cannot expect that a generic 9-component tensor can be represented by 4 components of the vector $v_i$. Thus (\ref{sym-1}) can be used only as an example of a special symmetric skewon.

In \cite{Obukhov:2004zz}, a  symmetric skewon parametrized by two vector fields was considered. 
\begin{equation}\label{sym-2}
S_{ij}=\frac 12 \left(u_iv_j+u_jv_i\right)-\frac 14 (u,v)g_{ij}\,.
\end{equation}
This tensor was treated as  a skewon with 8 independent components. In fact,  the rescaling of  the vector fields $u\to Cu, v\to C^{-1}v$ does not change the left hand side of Eq.(\ref{sym-2}), thus  the number of independent components is at least equal to 7.  
The form of the ansatz (\ref{sym-2}) can be modified by a reparametrization $u=k+\ell$, and  $ v=k-\ell$. This way we arrive  an  equivalent ansatz
\begin{equation}\label{sym-3}
S_{ij}= \left(k_ik_j-\frac 14 k^2g_{ij}\right)-\left(\ell_i\ell_j-\frac 14\ell^2g_{ij}\right)\,,
\end{equation}
that is more prone to generalization. 

The ansatz (\ref{sym-2}) includes less independent components than  needed for a generic symmetric skewon. Let us try to consider an ansatz constructed from three null vector fields $k,l,$ and $m$ and three real parameters $A,B,C$.
\begin{equation}\label{sym-4}
S_{ij}= Ak_ik_j+B\ell_i\ell_j+Cm_im_j\,,
\end{equation}
such that 
\begin{equation}\label{sym-5}
k^2=\ell^2=m^2=0\,.
\end{equation}
We observe that the tensor $S_{ij}$ in Eq.(\ref{sym-4}) is traceless. Moreover, every null vector can be considered as having 3 independent  components. Consequently, we can expect to have all together 9 independent components as it must be for a generic traceless symmetric matrix $S_{ij}$. 
Unfortunately, this skewon is not a most generic one. 
Indded, we immediately observe the determinant of the tensor (\ref{sym-4}) is equal to zero. In fact, for linearly independent vectors $k,\ell$, and $m$, the rank of the matrix  in Eq.(\ref{sym-4}) is equal to three. It means that the ansatz (\ref{sym-4}) is not most generic because it cannot   describe a matrix $S_{ij}$ of the  rank four. 

Consequently we come to our final vector ansatz
\begin{equation}\label{sym-7}
S_{ij}= Ak_ik_j+B\ell_i\ell_j+Cm_im_j+Dn_in_j\,,
\end{equation}
with 4 real parameters $A,B,C,D$ and 4 linearly independent null vectors $k,\ell,m,n$ such that 
\begin{equation}\label{sym-8}
{k^2=\ell^2=m^2=n^2=0}\,.
\end{equation}
Due to these relations, the matrix  $S_{ij}$ in Eq.(\ref{sym-7}) is traceless. 
For linearly independent vectors $k,\ell, m,n$, the determinant of the tensor in Eq.(\ref{sym-7}) is non-zero 
\begin{equation}\label{sym-7vol}
{\rm det}\, S_{ij}= ABCD\,({\rm vol})^2 \,,
\end{equation}
where {\rm vol} refers to the volume of the 4-dimensional parallelepiped determined by the vectors  $k,\ell,m,$ and $n$. 
Although the positive values of the coefficients $A,B,C,D$ can be absorbed into the vectors, it is convenient to preserve them in order to control the signs and the magnitude of the different contributions to the skewon field. 
 
We assume that the vectors $k,\ell,m,$ and $n$ are linearly independent. In the case of the non-zero coefficients $A,B,C,D$, the rank of the matrix $S_{ij}$ is equal to 4. When one of the coefficients is zero the rank is equal to 3, as in  ansatz (\ref{sym-4}). In the case, when two coefficients, say $C$ and $D$, equal to zero, we are left with a matrix of  rank  2. This is the case given in Eqs. (\ref{sym-2}) and  (\ref{sym-3}). When  three  coefficients, say $B,C, D,$, are zero,  the rank of the matrix $S_{ij}$ is equal to 1.  

Consequently, with our ansatz (\ref{sym-7}) we are able to describe a  symmetric skewon of  arbitrary rank. 
\subsection{Dispersion relation}
In order to derive the dispersion relation for the skewon (\ref{sym-7}), we start with the skewon covector. For the case of  generic gauge, it is given by
\begin{eqnarray}\label{sym-9}
Y_i&\!\!=\!\!&Ak_i(k,q)+B\ell_i(\ell,q)+Cm_i(m,q)+Dn_i(n,q)+\a q_i
\nonumber\\
&&
\end{eqnarray}
We choose the Lorenz-type gauge, $(Y,q)=0$ and derive the value of the parameter $\a$
\begin{eqnarray}\label{sym-10}
&&\!\!\!\!\!\!Y_i=A\frac{(k,q)}{q^2}\left(q^2k_i-(k,q)q_i\right)
\!+\!B\frac{(\ell,q)}{q^2}\left(q^2\ell_i-(\ell,q)q_i\right)+\nonumber\\
&& C\frac{(m,q)}{q^2}\left(q^2m_i-(m,q)q_i\right)+
D\frac{(n,q)}{q^2}\left(q^2n_i-(n,q)q_i\right)\nonumber\\&&
\end{eqnarray}
The square of this covector reads
\begin{eqnarray}\label{sym-11-cc}
&&\!\!\!\!Y^2=-\frac 1{q^2}
\left[A(k,q)^2\!+\!B(\ell,q)^2\!+\!C(m,q)^2\!+\!D(n,q)^2\right]^2+
\nonumber\\&&
\qquad 2\big[AB(k,q)(\ell,q)(k,\ell)+AC(k,q)(m,q)(k,m)+
\nonumber\\&&
\qquad \, \,\, \,  \,\, AD(k,q)(n,q)(k,n)+BC(\ell,q)(m,q)(\ell,n)+
\nonumber\\&&
\qquad \, \,\, \,  \,\, BD(\ell,q)(n,q)(\ell,n)+CD(m,q)(n,q)(m,n)\big]\,.
\nonumber\\&&
\end{eqnarray}
The dispersion relation $q^4=Y^2q^2$ now takes  the form
\begin{eqnarray}\label{sym-11}
q^4&\!\!=\!\!&-
\left[A(k,q)^2+B(\ell,q)^2+C(m,q)^2+D(n,q)^2\right]^2+
\nonumber\\&&
 2q^2\big[AB(k,q)(\ell,q)(k,\ell)+AC(k,q)(m,q)(k,m)+
\nonumber\\&&
 AD(k,q)(n,q)(k,n)+BC(\ell,q)(m,q)(\ell,n)+
\nonumber\\&&
 BD(\ell,q)(n,q)(\ell,n)+CD(m,q)(n,q)(m,n)\big]\,.
\nonumber\\&&
\end{eqnarray}
In order to study certain  conclusions of these formulas, we  consider  some specific examples. 
\subsection{Examples} 
\subsubsection{Skewon of rank 1} 
In the case $B=C=D=0$, we have a symmetric skewon of the lowest rank 1. The dispersion relation (\ref{sym-11}) takes the form
\begin{eqnarray}\label{sym-12}
q^4+A^2(k,q)^4=0
\end{eqnarray}
Since both terms in the left hand side  are positive, the unique solution must satisfy 
\begin{eqnarray}\label{sym-13}
q^2=0\,,\qquad (k,q)=0
\end{eqnarray}
Geometrically this is a low dimensional intersection of the light cone with a 3 dimensional hyperspace. It can be  a 
point $q_i=0$ or two second-dimensional lines lying on the light cone. It is quite doubtful whether  some physically meaningful situation  can be attributed to this solution. 
\subsubsection{Diagonal skewon of rank 2} 
A generic symmetric skewon of  rank 2 can be generated by choosing  $C=D=0$ in Eq.(\ref{sym-9}).
In this case, the dispersion relation takes the form
\begin{eqnarray}\label{sym-14}
q^4&=&-
\left[A(k,q)^2+B(\ell,q)^2\right]^2+
 2ABq^2(k,q)(\ell,q)(k,\ell)\,.
\nonumber\\&&
\end{eqnarray}
This relation can be considered as a quadratic equation relative to the variable $q^2$. In order to have real solutions, the following inequality must hold
\begin{equation}\label{sym-14a}
A^2B^2(k,q)^2(\ell,q)^2(k,\ell)^2\ge \left[A(k,q)^2+B(\ell,q)^2\right]^2\!\!.
\end{equation}

Let us consider, for example, a simplest symmetric  skewon of rank 2 with two non-zero components
\begin{equation}\label{sym-14b}
S_{00}=S_{11}=A\,.
\end{equation}
This skewon is evidently traceless, $g^{ij}S_{ij}=0$. 
It is convenient to proceed now with $Y_i=S_{ij}q^j$, i.e. in a gauge $\theta=0$ instead of the Lorenz-type gauge. 
The skewon covector takes the form 
\begin{equation}\label{sym-14b0}
Y_i=(A\omega,-Ak_1,0,0)\,.
\end{equation}
We calculate 
\begin{equation}\label{sym-14b1}
(Y,q)=A(\omega^2+k_1^2)\,, \qquad Y^2=A^2(\omega^2-k_1^2)\,.
\end{equation}
Thus the dispersion relation in Eq.(\ref{skew-16a})  takes the form 
\begin{equation}\label{sym-14bb}
q^4-A^2(\omega^2-{\rm k}_1^2)q^2+A^2(\omega^2+{\rm k}_1^2)^2=0\,.
\end{equation}
From this equation, we derive that the solution cannot be null. 
Moreover, every solution $q_i=(\omega, {\rm k}_1, {\rm k}_2, {\rm k}_3)$ must satisfy the inequality $|\omega|<|{\rm k}_1|$. It is with the correspondence with the inequality $q^2<0$. 

The solution of Eq.(\ref{sym-14bb}) can be rewritten as   
\begin{equation}\label{sym-15}
q^2=\frac{A^2}2\left(\omega^2-{\rm k}_1^2 \right) \pm \frac{A}2\sqrt{\left(A^2-4\right)\left(\omega^2-{\rm k}_1^2\right)^2-16\omega^2k_1^2 }
\end{equation}
Let us analyze the solutions of these equations for different values of the skewon parameter.
\begin{itemize}
\item [(1)] For $A=0$, we have here  the non-modified light cone $q^2=0$. 
\item [(2)] For $0<|A|\le 2$ there are no real solutions at all. 
\item [(3)] For $|A|> 2$, there  are two real solutions. 
\end{itemize}

 In Fig. 3 and Fig. 4 we depict  the images of these algebraic cones, 
We observe that the light cones intersect only at the origin, thus optic axes absent. 
  \begin{figure}[h!]
{
\includegraphics[width=6.0cm]{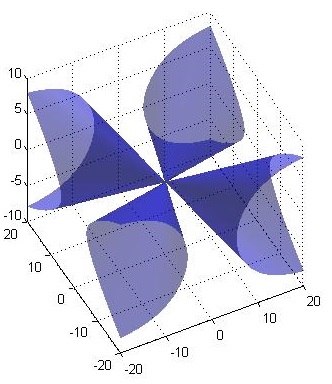}}
\caption[]{Section $k_3=0$ of the algebraic cones of the  diagonal rank 2 symmetric skewon (\ref{sym-14bb}).  The parameter $A=4$. \\ $w$ is directed as $z$-axis, $k_1,k_2$ are directed as $x$ and $y$ axes respectively.   }
\end{figure}
  \begin{figure}[h!]
{
\includegraphics[width=6.0cm]{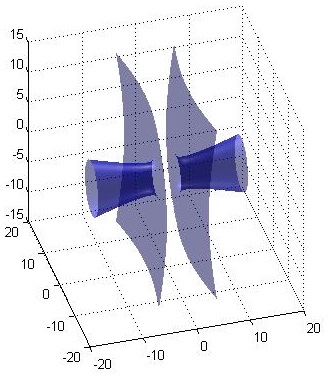}}
\caption[]{Section $w=1$ of the algebraic cones of the diagonal rank 2 symmetric skewon (\ref{sym-14bb}).  The parameters $A=4$.\\ $k_1,k_2,k_3$ are directed as $x,y$ and $z$ axes respectively.  }
\end{figure}
\subsubsection{Non-diagonal skewon of rank 2} 
We consider another example of a symmetric skewon of rank 2 of a non-diagonal type. Let the non-zero components of the symmetric traceless matrix $S_{ij}$ be
\begin{equation}\label{nond-1}
S_{01}= S_{10}=B\,.
\end{equation}
We assume the gauge $\theta=0$, consequently the skewon  covector takes the form $Y_i=S_{ij}q^j$. 
Its components are expressed as 
 \begin{equation}\label{nond-2}
Y_0=-B{\rm k}_1\,,\quad Y_1=Bw\,, \quad Y_2=Y_3=0 \,.
\end{equation}
Consequently,
 \begin{equation}\label{nond-3-cc}
Y^2=-B^2(\omega^2 -{\rm k}_1^2)\,,\qquad (Y,q)=-2Bw{\rm k}_1\,.
\end{equation}
The dispersion relation, $q^4=q^2Y^2-(q,Y)^2$ now becomes
 \begin{equation}\label{nond-3}
q^4+q^2B^2(\omega^2 -{\rm k}_1^2)+4B^2\omega^2{\rm k}_1^2=0\,.
\end{equation}
The solutions of this equation read
 \begin{equation}\label{nond-4}
q^2=-\frac{B^2}2(\omega^2 -{\rm k}_1^2)\pm \frac B2 \sqrt{B^2(\omega^2 -{\rm k}_1^2)^2-16\omega^2{\rm k}_1^2}\,.
\end{equation}
We observe that this solution exists for an arbitrary value of the parameter $B$. This means that the parametric gap is absent in this model.

In Fig. 5 and Fig. 6, we present the sections  light cones corresponding to (\ref{nond-3}). The cones do not intersect and  thus there are no the optic axes  in this case. 
\begin{figure}[h!]
{
\includegraphics[width=6.0cm]{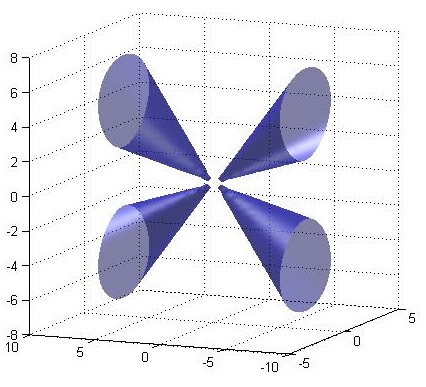}}
\caption[]{Section $k_3=0$ of the algebraic cones of the non-diagonal rank 2 symmetric skewon (\ref{sym-14bb}).  The parameter $B=4$. The component $w$ is directed as $z$-axis, $k_1,k_2$ are directed as $x$ and $y$ axes respectively.   }
\end{figure}
  \begin{figure}[h!]
{
\includegraphics[width=6.0cm]{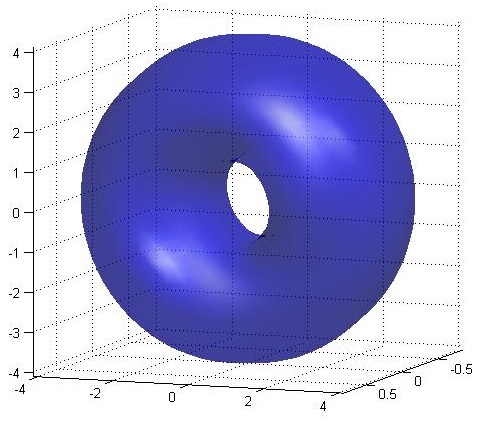}}
\caption[]{Section $w=1$ of the algebraic cones of the non-diagonal rank 2 symmetric skewon (\ref{sym-14bb}).  The parameters $B=4$. The components $k_1,k_2,k_3$ are directed as $x,y$ and $z$ axes respectively.  }
\end{figure}

\subsubsection{Skewon of rank 3} 
A generic symmetric skewon  of the rank 3 can be derived from Eq.(\ref{sym-9})  by choosing the parameter $D=0$. 
Instead, we consider a simplest diagonal symmetric traceless skewon of rank 3 with three non-zero components
\begin{equation}\label{sym-14c}
S_{00}=2A\,, \quad S_{11}=S_{22}=A\,.
\end{equation}
The non-zero components of the skewon covectror are
\begin{equation}\label{sym-20}
Y_0=2Aw,\quad Y_1=-Ak_1,\quad Y_2=-Ak_2\,.
\end{equation}
Consequently
\begin{equation}\label{sym-21}
(Y,q)=A\left(2\omega^2+z^2\right),\qquad 
Y^2=A^2\left(4\omega^2-z^2\right),
\end{equation}
where the notation $z^2=k_1^2+k_2^2$ is used.
The dispersion relation taken in the form $q^4-q^2Y^2+(Y,q)^2=0$ reads
\begin{equation}\label{sym-23}
q^4-A^2\left(4\omega^2-z^2\right)q^2 +A^2\left(2\omega^2+z^2\right)^2=0\,.
\end{equation} 
The solution of this biquadratic equation is
\begin{equation}\label{sym-24}
q^2\!=\!\frac {A^2}2\left(4\omega^2\!-z^2\right)\pm 
\frac A2\sqrt{A^2\left(4\omega^2\!-z^2\right)^2\!-4\left(2\omega^2+z^2\right)^2}\,.
\end{equation}
Rearranging the terms under the square root we get 
\begin{equation}\label{sym-25}
q^2=\frac {A^2}2\left(4\omega^2-z^2\right)\pm 
\frac A2\sqrt{f_1(\omega,z)f_2(\omega,z)}\,,
\end{equation}
where
\begin{equation}\label{sym-26}
f_1(\omega,z)=4(A-1)\omega^2-(A+2)z^2,\end{equation}
and
\begin{equation}\label{sym-27}
f_2(\omega,z)=4(A+1)\omega^2-(A-2)z^2\,.
\end{equation}

For $A=0$, Eq.(\ref{sym-25}) 
 represents the non-modified light cone $q^2=0$. 

For $0<|A|\le 1$, the term $f_1(w,z)$ is non-positive and the term $f_2(w,z)$ is non-negative. Moreover, if one of these terms vanishes, the unique solution  of Eq.(\ref{sym-23}) is the zero vector $q_i=0$. 
Consequently,  there are no non-zero solutions of the dispersion relation in this case.

For $1<|A|\le 2$, the discriminant  can be positive in some directions. However, the whole equation does not have real solutions. 
This can be easily observed taking into account the fact that $q^2\le 0$. 

\begin{figure}[h!]
{
\includegraphics[width=6.0cm]{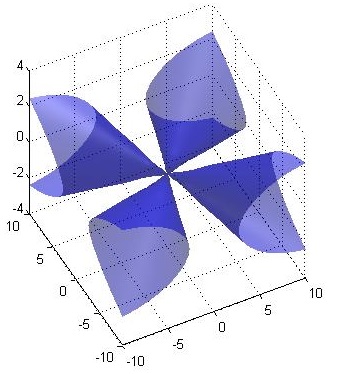}}
\caption[]{Section $k_2=0$ of the algebraic cones of the non-diagonal rank 2 symmetric skewon (\ref{sym-14bb}).  The parameter $A=4$. The component $w$ is directed as $z$-axis, $k_1,k_3$ are directed as $x$ and $y$ axes respectively.   }
\end{figure}

\begin{figure}[h!]
{
\includegraphics[width=6.0cm]{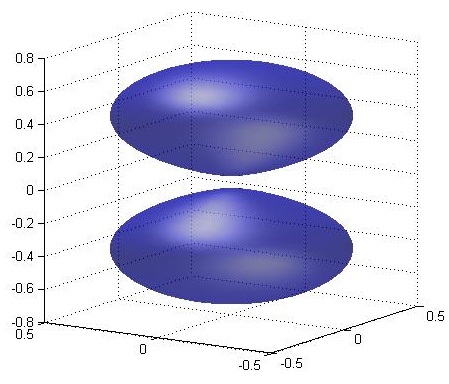}}
\caption[]{Section $w=0$ of the algebraic cones of the non-diagonal rank 2 symmetric skewon (\ref{sym-14bb}).  The parameter $A=10$. The component  $k_1,k_2,k_3$ are directed as $x,y$ and $z$ axes respectively.   }
\end{figure}

\subsubsection{Skewon of rank 4 }
In order to complete our analysis, we recall the known example of the diagonal  symmetric skewon of  rank 4. In the  form used in \cite{Obukhov:2004zz}, the non-zero components of its matrix $S_{ij}$ read
\begin{equation}\label{diag-1}
S_{00}=3A\,, \qquad S_{11}=S_{22}=S_{33}=A\,.
\end{equation}
We will use  the  gauge $\theta=0$, so the  skewon covector is given by $Y_i=S_{ij}q^j$. It   has the components ($\a=1,2,3$)
\begin{equation}\label{diag-2}
Y_0=3A w\,,\qquad Y_\a=-A{\rm k}_\a\,.
\end{equation}
Consequently, 
\begin{equation}\label{diag-2x}
Y^2=A^2(9\omega^2- {\rm k}^2)\,, \qquad (Y,q)=A(3\omega+{\rm k}^2)\,.
\end{equation}
The dispersion relation (\ref{skew-14}) now  takes the form
\begin{equation}\label{diag-3}
\omega^4-2\omega^2{\rm k}^2\left(1-8A^2\right)+{\rm k}^4=0\,.
\end{equation}
The solutions are  
\begin{equation}\label{diag-4}
\frac{\omega^2}{{\rm k}^2}=\left(1-8A^2\right)\pm 4\sqrt{4A^4-A^2}\,.
\end{equation}
Consequently we have the following cases:
\begin{itemize}
\item[(1)] For $|A|=0$, Eq.(\ref{diag-4}) has a unique solution -- the ordinary light cone ${\omega^2}={{\rm k}^2}$.
\item[(2)] For $0<|A|\le 1/2$, Eq.(\ref{diag-4}) does not have real solutions at all.
\item[(3)] For $1/2<|A|$, the right hand side of Eq.(\ref{diag-4}) is real and negative. Thus there is no wave propagation absents also in this case. 
\end{itemize}
Consequently, in the isotropic symmetric skewon model  wave propagation is absent.  This already was observed in \cite{Obukhov:2004zz}.


\section{Conclusions and discussion}
In this paper, we study an electromagnetic system with a generic local linear response. 
We presented here a new  formalism based on the notion of  optic tensor. This tensor completely describes the wave propagation; in particular, the dispersion relation is represented as the adjoint of the optic tensor. The decomposition of the optic tensor into symmetric and skew symmetric parts is directly related to the irreducible decomposition of the constitutive tensor into principle  and skewon parts, respectively. Moreover, we show that the skewon optic tensor is represented by the skewon covector, which is defined only up to addition of a term proportional to the wave covector. Consequently, the Lorenz-type gauge condition is introduced. These notations  simplify the dispersion relation considerably. 

We apply the optic tensor formalism to the problem of wave propagation in  media with a pseudo-Riemannian principle part and a generic skewon part. We proved  that independently of a specific type of the skewon field, the wave covector in such a medium is necessary space-like or null. 

Special types of the antisymmetric and symmetric skewon are considered. For the  antisymmetric skewon, the light cone  separates into two light cones -- the ordinary one $q^2=0$ and the extraordinary with $q^2>0$. This is a case of birefringence. We derived a degenerate case where the extraordinary light cone turns into a hyperplane. Moreover, for a special magnetic type, the extraordinary light cone has a space-like axis. This behavior can be interpreted as the situation where two cones are not observable both together, i.e., as absence of the birefringence phenomenon. For  all types of the  antisymmetric skewon, we proved  that there are exactly two optic axes and derive their explicit form.
 
Using the skewon covector formalism, we studied some new properties of  the symmetric skewon models. In particular we observed the parametric gap that  appears in most models, but not  in all of them.   In order to distinguish the  different types of the symmetric skewon we used the notion of rank of the corresponding matrix. This way we derived the general   vector  parametrization of the symmetric skewon. Dispersion relations for the symmetric skewon models of different rank where derived. We worked out some specific examples that present different types of  modified light cone structures.   

Our analysis shows that a skewon field is able to describe   certain new types of Lorentz violations in electrodynamics that are not accounted for in models based on a modified Lagrangian. 

In our opinion, the skewon model provides a rich subject  for algebraical analysis  and the physical considerations of the modified light cone structure. 

It is straightforward to observe that  the unusual features of  wave propagation  in the skewon media  are mostly related to  the real skewon. 
The situation changes crucially  when we turn to an imaginary skewon field. Such a case might be even more relevant in the problems of crystal physics. The corresponding treatment will be presented in a separate publication.  

\section*{Acknowledgments}
My acknowledgments to F. Hehl (Cologne/Missouri), Yu. Obukhov (Cologne/Moskow), V. Perlick (ZARM, Bremen), C. Laemmerzahl (ZARN, Bremen),  Y. Friedman (JCT, Jerusalem), and I. Godin (HUJI, Jerusalem) 
for valuable discussions.  
I acknowledge  the GIF grant No. 1078-107.14/2009 for financial support.

\appendix

\section{Calculation of the second adjoint}
For an arbitrary matrix $M^{ij}$, the second adjoint is defined as
\begin{equation}\label{apen-1}  
B_{ijkl}=\frac 1{2!}\ep_{ijmn}\ep_{klrs}M^{mr}
M^{ns}\,.
\end{equation}
Let us calculate this expression in the term of the skewon optic covector. Substituting  (\ref{chis-11}) into (\ref{apen-1}) we  write it as 
\begin{equation}\label{apen-2}  
B_{ijkl}(Q)=\frac 1{2!}\big(\ep_{ijmn}\ep^{mrab}q_aY_b\big)\cdot 
\big(\ep_{klrs}\ep^{nscd}q_cY_d\big)\,.
\end{equation}
Using the standard formulas for the product of two permutation tensors we have
\begin{equation}\label{apen-3-cc} 
 \ep_{ijmn}\ep^{mrab}q_aY_b=\left| \begin{array}{ccc}
\d^{r}_{i} &\d^{a}_{i}&\d^{b}_{i} \\
\d^{r}_{j} &\d^{a}_{j}&\d^{b}_{j} \\
\d^{r}_{n} &\d^{a}_{n}&\d^{b}_{n} 
\end{array} \right|q_aY_b=\left| \begin{array}{ccc}
\d^{r}_{i} &q_i&Y_i \\
\d^{r}_{j} &q_j&Y_j \\
\d^{r}_{n} &q_n&Y_n
\end{array} \right|\,.\end{equation}
Similarly,
\begin{equation}\label{apen-3} 
\ep_{klrs}\ep^{nscd}q_cY_d=\left| \begin{array}{ccc}
\d^{n}_{k} &\d^{c}_{k}&\d^{d}_{k} \\
\d^{n}_{l} &\d^{c}_{l}&\d^{d}_{l} \\
\d^{n}_{r} &\d^{c}_{r}&\d^{d}_{r}
\end{array} \right|q_cY_d=\left| \begin{array}{ccc}
\d^{r}_{i} &q_k&Y_k \\
\d^{r}_{j} &q_l&Y_l \\
\d^{r}_{n} &q_r&Y_r
\end{array} \right|\,.\end{equation}
Thus
\begin{eqnarray}
B_{ijkl}&=&\frac 12 \left| \begin{array}{ccc}
\d^{r}_{i} &q_i&Y_i \\
\d^{r}_{j} &q_j&Y_j \\
\d^{r}_{n} &q_n&Y_n
\end{array} \right|\cdot
\left| \begin{array}{ccc}
\d^{r}_{i} &q_k&Y_k \\
\d^{r}_{j} &q_l&Y_l \\
\d^{r}_{n} &q_r&Y_r
\end{array} \right|
\end{eqnarray}
Expanding the third order determinants we have
\begin{eqnarray}
B_{ijkl}&=&
\frac 12 \left(\d^r_i\left| \begin{array}{cc}
q_j &Y_j \\
q_n &Y_n 
\end{array} \right|-
\d^r_j\left| \begin{array}{cc}
q_i &Y_i \\
q_n &Y_n 
\end{array} \right|+
\d^r_n\left| \begin{array}{cc}
q_i &Y_i \\
q_j &Y_j 
\end{array} \right|
\right)\nonumber\\&&\cdot
\left(\d^n_k\left| \begin{array}{cc}
q_l &Y_l \\
q_r &Y_r 
\end{array} \right|-
\d^n_l\left| \begin{array}{cc}
q_k &Y_k \\
q_r &Y_r 
\end{array} \right|+
\d^n_r\left| \begin{array}{cc}
q_k &Y_k \\
q_l &Y_l 
\end{array} \right|
\right)\nonumber\\
\end{eqnarray}
Term by term multiplication yields
\begin{eqnarray}
B_{ijkl}&=&
\left| \begin{array}{cc}q_j &Y_j \\q_k &Y_k \end{array} \right|
\left| \begin{array}{cc}q_l &Y_l \\q_i &Y_i \end{array} \right|-
\left| \begin{array}{cc}q_j &Y_j \\q_l &Y_l \end{array} \right|
\left| \begin{array}{cc}q_k &Y_k \\q_i &Y_i \end{array} \right|\nonumber\\
&=&(q_jY_k-q_kY_j)(q_lY_i-q_iY_l)-\nonumber\\&&(q_jY_l-q_lY_j)(q_kY_i-q_iY_k)\nonumber\\
&=&(q_iY_j-q_jY_i)(q_kY_l-q_lY_k)\,.
\end{eqnarray}
Expanding these expressions we come to a compact formula
\begin{eqnarray}
B_{ijkl}&=&(q_iY_j-q_jY_i)(q_kY_l-q_lY_k)\,.
\end{eqnarray}
It can be also written in  matrix form
\begin{eqnarray}
B_{ijkl}&=&
\left| \begin{array}{cc}q_i &Y_i \\q_j &Y_j \end{array} \right|
\left| \begin{array}{cc}q_k &Y_k \\q_l &Y_l \end{array} \right|\,.
\end{eqnarray}
Note that the symmetry properties 
\begin{equation}\label{app-15}  
B_{ijkl}=-B_{jikl}=-B_{klij} \,
\end{equation}
and also
\begin{equation}\label{app-16}  
B_{ijkl}=B_{klij} \,
\end{equation}
evidently hold for these representations.
\section{Proof of Proposition 1}

It is useful to involve the  second adjoint notion. For a square matrix $M$, the second adjoint $ {}^{(2)}{\rm adj}(M)$ is defined by removing two rows and two columns. So it is represented as a 4-th order tensor 
\begin{equation}\label{disp-14}  
B_{ijkl}=\frac 1{2!}\ep_{ijmn}\ep_{klrs}M^{mr}
M^{ns}\,.
\end{equation}
Evidently,
\begin{equation}\label{disp-15}  
B_{ijkl}=-B_{jikl}=-B_{klij} \,.
\end{equation}
For a symmetric or a skew-symmetric matrix $M$, but not for a generic matrix,
\begin{equation}\label{disp-16}  
B_{ijkl}=B_{klij} \,.
\end{equation}
In term of the second adjoint, the dispersion relation (\ref{disp-12}) takes the form 
\begin{equation}\label{disp-17}  
A_{ik}(P) +B_{ijkl}(Q)P^{jl}=0 \,,
\end{equation}
 In our formalism, the skewon  contribution to the dispersion relation is completely represented by the second term of (\ref{disp-17}).  
Let us calculate this expression in the term of the skewon optic covector. Substituting  (\ref{chis-11}) into (\ref{disp-14}) we  have 
\begin{equation}\label{disp-19}  
B_{ijkl}(Q)=\frac 1{2!}\ep_{ijmn}\ep_{klrs}\ep^{mrab}
\ep^{nscd}q_aq_cY_bY_d\,.
\end{equation}
Using the standard formulas for the product of permutation tensors, we calculate
\begin{equation}\label{disp-20}  
\ep_{ijmn}\ep^{mrab}q_aY_b=\left| \begin{array}{ccc}
\d^{r}_{i} &q_i&Y_i \\
\d^{r}_{j} &q_j&Y_j \\
\d^{r}_{n} &q_n&Y_n
\end{array} \right|\,.\end{equation}
Consequently, 
\begin{equation}\label{disp-21}  
B_{ijkl}(Q)=\frac 1{2!}\left| \begin{array}{ccc}
\d^{r}_{i} &q_i&Y_i \\
\d^{r}_{j} &q_j&Y_j \\
\d^{r}_{n} &q_n&Y_n
\end{array} \right|\,\cdot\,
\left| \begin{array}{ccc}
\d^{n}_{k} &q_k&Y_k \\
\d^{n}_{l} &q_l&Y_l \\
\d^{n}_{r} &q_r&Y_r
\end{array} \right|\,.
\end{equation}
Evaluating the determinants we obtain a compact formula
\begin{equation}\label{disp-22}  
B_{ijkl}(Q)=\left| \begin{array}{cc}
q_i&Y_i \\
q_j&Y_j 
\end{array} \right|\,\,\cdot\,
\left| \begin{array}{cc}
q_k&Y_k \\
q_l&Y_l 
\end{array}\right|\,.
\end{equation}
Consequently, the contribution of the skewon part into the dispersion relation is given by
\begin{eqnarray}\label{disp-23}  
P^{jl}B_{ijkl}(Q)&=&P^{jl}\left| \begin{array}{cc}
q_i&Y_i \\
q_j&Y_j 
\end{array} \right|\,\,\cdot\,
\left| \begin{array}{cc}
q_k&Y_k \\
q_l&Y_l 
\end{array}\right|\nonumber\\&=&q_iq_k\left(P^{jl}Y_jY_l\right)\,.
\end{eqnarray}
Observe that this expression is given by a scalar multiplied by $q_iq_k$. It is in a correspondence with (\ref{Adj2}). Thus we proved Proposition 1.

\end{document}